# Nanoparticle radiosensitization: a mass-energy-transfer extension of the compound Poisson killing model to high-Z nanoparticle-mediated local-dose and radiolysis enhancement

Xinglan Qin[1,2], Zixu Zhao[1,2], Xiaolong Sha[1,2], Zhaoxin Yang[1,2], and Xufei Wang[1,2 *]

1. Institute of Modern Physics, Fudan University, Shanghai 200433, China
2. Key Laboratory of Nuclear Physics and Ion-beam Application (MOE), Fudan University, Shanghai 200433, China

**E-mail:** wangxufei@fudan.edu.cn
**Keywords:** high-Z nanoparticles, radiosensitization, compound Poisson killing model, nanoparticle agglomeration, mass-energy transfer, secondary-electron emission

## Abstract

*Objective.* To quantify the biological effectiveness of high-Z nanoparticle radiosensitization, the compound Poisson killing (CPK) model, originally for nanoparticle surface-mediated biochemical enhancement, is extended into a unified framework integrating both surface biochemical enhancement and mass-energy-transfer-mediated local-dose and radiolysis enhancement. *Approach.* Retaining the core compound Poisson additive killing assumption and per-nanoparticle equivalent dose definition, nanoparticle-induced cell killing for both the enhancement pathways is described under the linear quadratic (LQ) model. The agglomeration dependence of the equivalent doses is mapped onto equal-mass isolated single-sphere radii via two mappings: matched effective area and matched energy emission ratio. Two phenomenological factors $t$ and $\eta$ encode agglomeration effects, yielding the separate CPK-I (surface biochemical) and CPK-II (mass-energy-transfer) models for the LQ parameter modification α''/α. A generalized dual enhancement formulation is thus constructed. To reduce free parameters, the $t$-$\eta$ correlation is analyzed. Monte Carlo simulations reveal an inherent inequality $R_\beta > R_S$ between the area-equivalent radius $R_S$ and energy-transfer-ratio-equivalent radius $R_\beta$, imposing a constraint $t > 1-\eta$. This inequality arises from the lower energy emission ratio of agglomerated nanoparticle systems relative to their equal-area single-sphere counterparts, yielding a nonlinear $t$-$(1-\eta)$ relationship and a parameter-reduced expression of α''/α that accounts for both enhancements. The general α''/α expression gives three prediction models: CPK-iso, CPK-I, and CPK-II. Interpolative validation was performed using the clonogenic dose-survival data from HepG2 cells exposed to polyethylene glycol-coated gold nanoparticles (PEG-GNP). *Results.* Statistical evaluation indicates that the CPK-iso model achieves the highest accuracy and stability with the errors within ±5%. CPK-II performs well with negligible bias, while CPK-I yields larger deviations and concentration-dependent overestimation for PEG-GNP radiosensitization dominated by secondary-electron local dose and radiolysis enhancement. *Significance.* This extended CPK model decouples mass-energy-transfer-mediated enhancement from surface-mediated biochemical enhancement, enabling mechanism-driven quantitative prediction of clonogenic dose-survival response for high-Z nanoparticle radiosensitization.

## 1. Introduction

Over the past two decades, nanoparticle radiosensitization has been intensively investigated as an effective strategy to boost radiotherapy performance (Abulimiti et al 2025, Carbone et al 2025, Cizkova et al 2024, Schick et al 2024, Penninckx et al 2020, Schuemann et al 2020, Sajo and Zygmanski 2020). Substantial progress has been made via *in silico, in vitro* and *in vivo* studies, ranging from high-Z metallic nanoparticle sensitizers (Loscertales et al 2025, Colak and Ertas 2024, Song et al 2022, Kempson 2021, Yang et al 2020, Zhang et al 2021) to "sensitizer-plus" nanoagents designed for multifunctional theranostic applications (Pan et al 2024, Lin et al 2025, Wedler et al 2025, Guo et al 2025), alongside a deeper understanding of the radiosensitization mechanisms, including microdosimetric enhancement (Fardirad et al 2025, Polopetrakis et al 2025, Thomas et al 2024, 2026, Ohno et al 2021), physicochemical-to-biochemical enhancements (Fardirad et al 2025, Antunes et al 2025, Zheng and Sanche 2023, Howard et al 2020, Peukert et al 2019, 2020, Klein et al 2014, 2018, Butterworth et al 2012, 2013), and biological mechanisms including cell cycle arrest, endoplasmic reticulum stress, DNA repair inhibition, and ferroptosis (Sun et al 2020, Tremi et al 2021, 2022, Hu et al 2024, Liu et al 2018, Sun et al 2022, Darmon et al 2022). Beyond photon radiotherapy, nanoparticle radiosensitization for proton and heavy-ion beams has also attracted increasing attention (Carbone et al 2025, Ma et al 2024, Gerken et al 2022, Lo et al 2023, Rudek et al 2019, Feng et al 2018), given the favorable energy deposition features alongside the inherent limitation in biological effectiveness, especially for protons. Meanwhile, for preclinical planning of nanoparticle-enhanced radiotherapy, advances have also been achieved in biological effectiveness modeling (McMahon et al 2011, 2016, Lechtman et al 2013, Zygmanski et al 2013, Lin et al 2014, 2015, Lechtman and Pignol 2017, Ferrero et al 2017, Sung et al 2017, 2018, Sung and Schuemann 2018, Brown and Currell 2017, Villagomez-Bernabe and Currell 2019, Villagomez-Bernabe et al 2021, Kim et al 2021, Engels et al 2020, Batooei et al 2021, Hahn and Zutta Villate 2021, Melo-Bernal et al 2018, 2021, Pan et al 2022, Schmidt et al 2022, Velten and Tomé 2023, Khaledi et al 2024, Rabus and Thomas 2025). Based on these progresses, clinical trials also yielded promising results (Chargari et al 2024, Dubey et al 2022, Scher et al 2020, Bilynsky et al 2021, Bonvalot et al 2019, Verry et al 2019, 2021, Hoffmann et al 2021, Bort et al 2020).

Despite these encouraging advances, a comprehensive dosimetric model for nanoparticle-mediated biological effectiveness is still lacking, owing to the challenges in modeling the highly coupled, extremely inhomogeneous subcellular enhancement effects. Most existing biological-effectiveness models adopt local-dose assumptions, in which Monte Carlo simulations of secondary-electron deposition from high-Z metallic nanoparticles typically assume the nucleus to be the only radiosensitive target within a cell. Such predictions are strongly dependent on predefined subcellular source–target geometries and ignore local-dose-mediated killing originating from cytoplasmic organelles, as well as local-to-global biochemical effects across the entire target cell. To enable quantification of whole-cell killing events originating from both the nucleus and cytoplasm and incorporating both local-dose and local-to-global biochemical effects for a complete prediction of survival modification, fundamental challenges remain unresolved. First, quantitative modeling of non-nuclear-targeted killing is limited by poorly characterized cytoplasmic dose–response relationships. A second limitation is that few models are capable of quantifying cell killing arising from nanoparticle-mediated local-to-global biochemical effects, although biochemical contributions to the radiosensitization have been widely demonstrated for various nanoparticles.

To circumvent these difficulties, our previous work (Pan et al 2022) proposed a survival modification framework based on compound Poisson additive killing as a statistical alternative to local-effect modeling. This framework treats nanoparticle-mediated survival modification as a consequence of independent additive killing, where the assumed equivalent doses per nanoparticle are independently imposed on a cell under the LQ model. Nevertheless, the original compound Poisson killing (CPK) model for nanoparticle radiosensitization, built upon the assumption of surface-mediated enhancement, is primarily valid for surface-mediated biochemical enhancement induced by low-Z nanoparticles. It is not self-consistently applicable to mass-energy-transfer-mediated local-dose and radiolysis enhancement that arises from secondary-electron emission — the dominant physicochemical process induced by high-Z nanoparticles and often coupled with biochemical enhancement effects commonly triggered by intracellular nanomaterials.

To address this limitation, this study introduces a mass-energy-transfer extension of the original CPK model to incorporate the local-dose and radiolysis enhancement triggered by high-Z nanoparticles. The rest of this article is organised as follows. Section 2 lays out the theoretical framework of the extended CPK model. Section 2.1 introduces the baseline dose–survival model of intrinsic radiation cell killing. Section 2.2 establishes the general formulation of nanoparticle-mediated clonogenic survival modification under the assumption of compound Poisson additive killing, including its general framework and the derivation of key statistical and equivalent-dose quantities. Sections 2.3 and 2.4 construct CPK-I and CPK-II, respectively: CPK-I addresses the LQ-parameter modifications resulting from nanoparticle-surface-mediated biochemical enhancement, while CPK-II extends the model to incorporate mass-energy-transfer-mediated local-dose and radiolysis enhancement by high-Z nanoparticles. Section 2.5 further analyses the physical correlation between t and η — the two agglomeration factors independently introduced in CPK-I and CPK-II — to achieve parameter reduction and develop practical prediction models. Section 3 reports the experimental validation against HepG2 cell radiobiology data. Section 4 further discusses the physical origin behind the t–η correlation. Section 5 summarises the main conclusions. Supplementary details are given in Appendix.

## 2. Theoretical

### 2.1 Intrinsic dose-survival model for radiation-induced cell killing

In a single cell exposed to uniform radiation, the number of lethal damages (killing events) $l$ follows a binomial distribution that converges to a Poisson distribution as the incident photon or particle number for dose $D$ approaches infinity.

$$p(l|D) = \frac{\lambda(D)^l}{l!} e^{-\lambda(D)} \tag{1}$$

$\lambda(D)$ denotes the Poisson mean of intracellular killing events induced by a radiation dose of $D$. Accordingly, the survival probability corresponds to the probability of zero killing events within a single cell irradiated at dose $D$

$$S(D) = p(l{=}0|D) = \frac{\lambda(D)^0}{0!} e^{-\lambda(D)} = e^{-\lambda(D)} \tag{2}$$

For most cells irradiated with uniform radiation, the Poisson mean of intrinsic killing events per cell is described by the linear-quadratic (LQ) relationship

$$\lambda(D) = -\ln S(D) = \alpha D + \beta D^2 \tag{3}$$

Assuming that initial lethal lesions arising in the nucleus and cytoplasm are mutually independent, according to Raikov's theorem, the two components of killing events each follow the Poisson distribution

$$p_n(l|D) = \frac{\lambda_n(D)}{l!} e^{-\lambda_n(D)}, \quad p_{cy}(l|D) = \frac{\lambda_{cy}(D)}{l!} e^{-\lambda_{cy}(D)}, \quad \lambda_n(D) + \lambda_{cy}(D) = \lambda(D) \tag{4}$$

where $\lambda_n(D), \lambda_{cy}(D)$ are the Poisson means of killing events in nucleus and cytoplasm, respectively. For cytoplasm targets, lower radiosensitivity implies that dose-dependent lethal events across the whole cell are dominated by intranuclear killing mechanisms originating from lethal DNA damage via single- and double-strand breaks, so that equation (4) can be rewritten as

$$\begin{cases} p_n(l|D) = p(l|D) = \dfrac{(\alpha D + \beta D^2)^l}{l!} e^{-(\alpha D + \beta D^2)}, & \lambda_n(D) \approx \lambda(D) = \alpha D + \beta D^2 \\ p_{cy}(l|D) = \dfrac{\nu(D)^l}{l!} e^{-\nu(D)}, & \lambda_{cy}(D) = \nu(D) \ll \alpha D + \beta D^2 \end{cases} \tag{5}$$

Therefore, the single-cell dose-survival probability can be expressed in a generalized form as

$$S(D) = p_n(l{=}0|D) \times p_{cy}(l{=}0|D) = e^{-(\alpha D + \beta D^2)} \times e^{-\nu(D)} \approx e^{-(\alpha D + \beta D^2)} \tag{6}$$

### 2.2 Nanoparticle-mediated survival modification via compound Poisson additive killing

*2.2.1 Basic framework*

Assuming that intrinsic radiation killing and nanoparticle-mediated exogenous killing occur independently within a single cell, a general formula for nanoparticle-modulated cell survival can be written as

$$S(D, C) = P(l{=}0|D) \times P(L_Z{=}0|D, C) \approx e^{-(\alpha D + \beta D^2)} \gamma(D, C) \tag{7}$$

$P(l{=}0|D)$ denotes the intrinsic probability of zero killing events (i.e., survival) under radiation dose $D$. $L_Z$ represents the number of killing events added by nanoparticles, and $P(L_Z{=}0|D, C)$ is the probability of zero killing events mediated by nanoparticles at dose $D$ and treatment concentration $C$. The intrinsic survival probability $P(l{=}0|D)$ follows the LQ model as given in equation (6), while the probability of zero nanoparticle-mediated killing events is defined as a modification function acting on the intrinsic dose-survival response

$$\gamma(D, C) = P(L_Z{=}0|D, C) \tag{8}$$

According to the general statistical properties of nanoparticles within a single cell, two fundamental random quantities are assumed to follow mutually independent Poisson distributions: (1) the number of nanoparticles acting on each cell Z, which denotes the count of internalized and pericellular nanoparticles and satisfies $Z \sim Pois(\bar{\lambda})$; (2) the number of killing events induced by each nanoparticle $L_1$, i.e. the incremental killing events mediated by a single nanoparticle inside a cell, which follows the Poisson distribution $L_1 \sim Pois(\mu_1)$. The mean number of nanoparticles per cell $\bar{\lambda}$ is correlated with nanoparticle treatment concentration $C$, while the mean number of incremental killing events per nanoparticle $\mu_1$ depends on both the nanoparticle treatment concentration $C$ and the radiation dose $D$. The two Poisson variables are expressed as

$$\begin{cases} Z \sim Pois(\bar{\lambda}) \to P(Z{=}z) = \dfrac{\bar{\lambda}^z}{z!} e^{-\bar{\lambda}}, & E(Z) = V(Z) = \bar{\lambda} = \bar{\lambda}(C) \\ L_1 \sim Pois(\mu_1) \to P(L_1{=}l) = \dfrac{{\mu_1}^l}{l!} e^{-\mu_1}, & E(L_1) = V(L_1) = \mu_1 = \mu_1(D, C) \end{cases} \tag{9}$$

Based on these two Poisson distributions and the further assumption that killing events induced by individual nanoparticles are independent, the total number of killing events per cell constitutes an additive process governed by a compound Poisson distribution

$$P(L_Z = l|D,C) = P\left(\sum_{i=1}^{Z} L_{1i} = l\right) = \sum_{n=0}^{\infty} P\left(\sum_{i=1}^{n} L_{1i} = l\right) P(Z=n) = \sum_{n=0}^{\infty}\left[\left(\frac{\bar{\lambda}^n}{n!}e^{-\bar{\lambda}}\right) P\left(\sum_{i=1}^{n} L_{1i} = l\right)\right]$$
$$= \sum_{n=0}^{\infty}\left[\frac{(n\mu_1)^l}{l!}e^{-n\mu_1}\right]\left(\frac{\bar{\lambda}^n}{n!}e^{-\bar{\lambda}}\right), \quad \begin{pmatrix} E(L_Z) = \bar{\lambda}\mu_1 \\ V(L_Z) = \bar{\lambda}\mu_1(1+\mu_1) \\ \bar{\lambda} = \bar{\lambda}(C), \mu_1 = \mu_1(D,C) \end{pmatrix} \tag{10}$$

Thus, the probability of zero lethal events induced by all nanoparticles per cell (i.e., the survival modification function) is

$$P(L_Z{=}0|D,C) = \gamma(D,C) = \sum_{n=0}^{\infty} e^{-n\mu_1(D,C)}\left(\frac{\bar{\lambda}(C)^n}{n!}e^{-\bar{\lambda}(C)}\right) \tag{11}$$

Substituting equation (11) into (7) yields the modified cell survival probability under compound Poisson killing

$$S(D,C) = P(l{=}0|D) \times P(L_Z{=}0|D,C) = e^{-(\alpha D+\beta D^2)}\sum_{n=0}^{\infty} e^{-n\mu_1(D,C)}\left(\frac{\bar{\lambda}(C)^n}{n!}e^{-\bar{\lambda}(C)}\right) \tag{12}$$

*2.2.2 $\bar{\lambda}(C)$: reduction of compound Poisson killing to a Poisson distribution*

The number of nanoparticles per cell can be approximated by the deterministic mean value $Z{=}\bar{\lambda}(C)$, instead of the Poisson random variable $Z{\sim}Pois(\bar{\lambda})$. Consequently, the compound Poisson distribution describing the additive killing in equation (10) is reduced to a Poisson distribution

$$P(L_Z = l|D,C) = \frac{\left(\bar{\lambda}(C)\mu_1(D,C)\right)^l}{l!}e^{-\bar{\lambda}(C)\mu_1(D,C)} \tag{13}$$

Accordingly, the survival modification function is reduced from equation (11) to

$$P(L_Z = 0|D,C) = \gamma(D,C) = e^{-\bar{\lambda}(C)\mu_1(D,C)} \tag{14}$$

With this Poisson approximation, the cell survival probability under the compound Poisson killing thus reduces to

$$S(D,C) = e^{-(\alpha D+\beta D^2)}e^{-\bar{\lambda}(C)\mu_1(D,C)}. \tag{15}$$

To reflect the saturation behavior of cellular uptake of nanoparticles with increasing treatment concentration, an exponential saturation function is adopted to describe the nonlinear concentration dependence of $\bar{\lambda}(C)$ (Chithrani *et al*, 2006; Kalambur *et al.*, 2007; Tahara *et al.*, 2009; Jain *et al.*, 2011; Lunov *et al.*, 2011; Jiang *et al.*, 2013)

$$\bar{\lambda}(C) \approx A(1-e^{-kC}) \tag{16}$$

where $A$ denotes the saturated mean number of nanoparticles taken up per cell, $k$ is the rate constant with units of reciprocal concentration. The parameters $A$ and $k$ depend on cell type, nanoparticle properties, and co-incubation conditions.

*2.2.3. $\bar{d}_1(D,C)$: equivalent dose per nanoparticle for quantifying cell-killing events*

The Poisson mean of killing events per nanoparticle $\mu_1(D,C)$ also requires an explicit expression. For this, an hypothetical spatially uniform dose over the whole cell, termed the equivalent dose $\bar{d}_1$ per nanoparticle, is introduced to quantify the Poisson killing events $L_1{\sim}Pois(\mu_1(D,C))$ induced per nanoparticle. Therefore, under the LQ radiation-killing model, the Poisson mean $\mu_1$of killing events per nanoparticle can be formulated as an LQ function of the equivalent dose $\bar{d}_1(D,C)$

$$\mu_1(D,C) = \alpha\bar{d}_1(D,C) + \beta{\bar{d}_1}^2(D,C). \tag{17}$$

Consistent with the independence assumption for lethal events induced by distinct nanoparticles adopted in equation (10), the overall cell-killing effect of multiple nanoparticles can be described as additive contributions of individual equivalent doses within the LQ model, with quadratic cross-terms neglected. In other words, the incremental lethal events arising from the subcellular physical-biochemical effects of intracellular nanoparticles are equivalent to the additive radiation-killing effect produced by multiple independent irradiations at the uniform equivalent dose $\bar{d}_1$.

## 2.3 CPK-I: LQ parameter modification for surface-mediated biochemical enhancement

*2.3.1. $\bar{d}_1(D,\bar{\lambda}(C))$: expression derivation*

The bivariate dependency of $\mu_1(D,C)$ can be expressed more explicitly as $\mu_1(D,\bar{\lambda}(C))$. Accordingly, the equivalent dose per nanoparticle $\bar{d}_1(D,C)$ is rewritten as $\bar{d}_1(D,\bar{\lambda}(C))$. Given the typial features of biochemical radiosensitization mediated through exposed surface of randomly agglomerated nanoparticles, the equivalent dose $\bar{d}_1(D,\bar{\lambda}(C))$, defined to quantify the incremental killing events induced by each nanoparticle within a cell, should have an explicit dependence on the effective surface area $\bar{s}_1$of an individual nanoparticle. To determine the effective surface area $\bar{s}_1$ per nanoparticle, we first define the total effective area of all the nanoparticles per cell, denoted as $\bar{S}$, which represents the total exposed area of the nanoparticles responsible for mediating these collective biological effects. The effective surface area per nanoparticle $\bar{s}_1$is then obtained by dividing the total effective area $\bar{S}$ by the mean number of nanoparticles per cell $\bar{\lambda}(C)$. Uisng this defined per-nanoparticle effective area $\bar{s}_1$, the equivalent dose per nanoparticle $\bar{d}_1$ is assumed to depend linearly on both $\bar{s}_1$ and the radiation dose $D$, yielding the explicit formulation

$$\bar{d}_1(D,C) = \bar{d}_1(D,\bar{\lambda}(C)) = K_s D \bar{s}_1(\bar{\lambda}(C)) \tag{18}$$

where $K_s$ is the proportional constant between the equivalent dose $\bar{d}_1$ and the applied dose $D$ per unit effective surface area. The geometric dependencies of the total effective area $\bar{S}$, the effective area per nanoparticle $\bar{s}_1$, and the equivalent dose per nanoparticle $\bar{d}_1$ on the mean number of nanoparticles per cell $\bar{\lambda}$ are derived as follows: Let $s_0$ denote the intrinsic surface area of a single nanoparticle. For fully-isolated nanoparticles of negligible agglomeration, $\bar{S}(\bar{\lambda})=s_0\bar{\lambda}$, $\bar{s}_1(\bar{\lambda})=\bar{S}/\bar{\lambda}=s_0$, thus $\bar{d}_1=K_s D s_0$. In the limiting case of maximal agglomeration into a compact single sphere $\bar{S}(\bar{\lambda})=s_0\bar{\lambda}^{2/3}$, $\bar{s}_1(\bar{\lambda})=\bar{S}/\bar{\lambda}=s_0\bar{\lambda}^{-1/3}$, thus $\bar{d}_1=K_s D s_0\bar{\lambda}^{-1/3}$. By interpolating between these two limiting scenarios, the total effective area $\bar{S}$ and the effective area per nanoparticle $\bar{s}_1$ for naturally agglomerated nanoparticles can be formulated as $\bar{S}(\bar{\lambda})=s_0\bar{\lambda}^{\eta}$, $\bar{s}_1(\bar{\lambda})=\bar{S}/\bar{\lambda}=s_0\bar{\lambda}^{\eta-1}$, thus the generalised expression of the equivalent dose per nanoparticle $\bar{d}_1$ can be expressed as

$$\bar{d}_1(D,\bar{\lambda}(C))=K_s D s_0 \bar{\lambda}(C)^{\eta-1}, \quad \eta \in (2/3,1) \tag{19}$$

The exponent $\eta$ acts as a dispersion- or agglomeration-characterizing factor ranging from 2/3 to 1, where an increase in $\eta$ leads to greater nanoparticle dispersity or lower agglomeration extent. Further details regarding the dependency derivations for the correlated chain $\bar{\lambda}\rightarrow\bar{S}\rightarrow\bar{s}_1\rightarrow\bar{d}_1$ can be found in our previously work (Pan et al., 2022).

#### 2.3.2. LQ parameter modification: relative corrections $\alpha''/\alpha$ and $\beta''/\beta$

By substituting equation (19) (with $\bar{\lambda}(C)$ simplfied as $\bar{\lambda}$) into (17) and the result into (15), the survival modification under the Poisson approximation for incremental killing events per cell can be described as a consequence of independent additive LQ killing (with quadratic cross-terms omitted) by the equivalent doses of the individual intracellular nanoparticles

$$\begin{aligned} S(D,C) &= e^{-(\alpha D+\beta D^2)} e^{-\bar{\lambda}\left[\alpha(K_s D s_0 \bar{\lambda}^{\eta-1})+\beta(K_s D s_0\bar{\lambda}^{\eta-1})^2\right]} \\ &= e^{-(\alpha D+\beta D^2)} e^{-(\alpha'' D+\beta'' D^2)} \\ &= e^{-[(\alpha+\alpha'')D+(\beta+\beta'')D^2]} \end{aligned} \tag{20}$$

The apparent parameters $(\alpha'',\beta'')$ are correction terms for the intrinsic LQ parameters $(\alpha,\beta)$. They quantify the cell survival modification induced by nanoparticle-mediated additive killing events and scale linearly with the intrinsic $\alpha$ and $\beta$. The proportional coefficients $\alpha''/\alpha$ and $\beta''/\beta$ (i.e., relative corrections) exhibit a nonlinear dependence on the mean number of nanoparticles per cell $\bar{\lambda}$

$$\frac{\alpha''}{\alpha} = \bar{\lambda}\left(K_s s_0 \bar{\lambda}^{\eta-1}\right), \qquad \frac{\beta''}{\beta} = \bar{\lambda}(K_s s_0\bar{\lambda}^{\eta-1})^2 \tag{21}$$

#### 2.3.3. Simplification to predictive models

From $\bar{d}_1=K_s D s_0\bar{\lambda}^{\eta-1}$ in equation (19), the relative corrections $\alpha''/\alpha$ and $\beta''/\beta$ given by equation (21) can be rewritten as

$$\frac{\alpha''}{\alpha} = \bar{\lambda}\frac{\bar{d}_1}{D}, \qquad \frac{\beta''}{\beta} = \bar{\lambda}\frac{\bar{d}_1}{D}\frac{\bar{d}_1}{D} \tag{22}$$

Physically, the induction coefficient $\varphi_1=\bar{d}_1/D$ for the equivalent dose of a single nanoparticle is expected to be far smaller than unity $(\varphi_1 \ll 1)$. We may further assume that the cumulative contribution of $\varphi_1$ over the total number of nanoparticles per cell remains below unity, $\alpha''/\alpha=\bar{\lambda}\varphi_1<1$. Accordingly, $\beta''/\beta=(\alpha''/\alpha)\varphi_1 \ll \alpha''/\alpha <1$, indicating that this quadratic correction term is negligible in practice. By this simplification, the LQ parameter modification reduces to a correction solely on the $\alpha$ parameter

$$\frac{\alpha''}{\alpha} = K_s s_0 \bar{\lambda}^{\eta}, \quad \frac{\beta''}{\beta} \approx 0 \tag{23}$$

Substituting the expression for $\bar{\lambda}(C)$ (equation (16)) back into (23) yields the relative correction $\alpha''/\alpha$ as a three-parameter function of nanoparticle concentration $C$

$$\frac{\alpha''}{\alpha} = K_s s_0 [A(1-e^{-kC})]^{\eta} \tag{24}$$

where the three independent fitting parameters are $A_S=K_s s_0 A^{\eta}$, $k$ and $\eta$. This formulation can be recast as a two-parameter expression for $(\alpha''/\alpha)^{1/\eta}$ in terms of concentration $C$

$$\left(\frac{\alpha''}{\alpha}\right)^{\frac{1}{\eta}} = A'(1-e^{-kC}), \qquad (2/3<\eta \le 1) \tag{25}$$

For a fixed $\eta$, the two unknown parameters $A'=(K_s s_0)^{1/\eta} A$ and $k$ can be directly determined by fitting $(\alpha''/\alpha)^{1/\eta}$ against concentration $C$ using two pairs of $(\alpha''/\alpha, C)$ datasets measured at $C_1$ and $C_2$. The fitted paramneters are subsequently adopted to interpolate the value of $(\alpha''/\alpha)^{1/\eta}$, from which the relative correction $\alpha''/\alpha$ can be predicted for any intermediate concentration satisfying $C_1<C<C_2$.

In practice, the interpolative prediction of $\alpha''/\alpha$ can be carried out via two alternative models. For the fully isolated model, the unkown parameters $(A',k)$ are determined from equation (25) using two pairs of measured $(\alpha''/\alpha, C)$ datasets at $\eta=1$. The fitted parameters are then directly used to interpolate the $\alpha''/\alpha$ value corresponding to any intermediate concentration satisfying $C_1<C<C_2$. In the agglomeration model, the parameters $(A', k)$ are calculated from equation (25) using the same set of $(\alpha''/\alpha, C)$ datasets, yet yield distinct parameter solutions for different values of $\eta\in(2/3, 1)$. The optimal $\eta$, together with the corresponding $(A', k)$, is determined by fitting against experimental data acquired at one or more intermediate concentrations. These calibrated parameters are subsequently employed to produce more accurate predictions of $\alpha''/\alpha$ at other intermediate concentrations.

#### *2.3.4. Limitations*

*In vitro* validation using carbon-dot-exposed HepG2 cells (Pan et al., 2022) verified the high accuracy of the CPK-I model for predicting LQ-parameter modification to the dose-survival relationship. However, when applied to high-Z nanoparticles, the underlying assumption adopted in equation (18) that the equivalent dose $\bar{d}_1$ of an individual nanoparticle varies linearly with its effective surface area $\bar{s}_1$carries inherent limitations, becasue the local-dose and radiolysis enhancement driven by secondary-electron-mediated mass-energy transfer is a physical process that does not scale simply with the exposed surface area of intracellular nanoparticles. Therefore an extended framework is needed to formulate the per-nanoparticle equivalent dose $\bar{d}_1$ for individual high-Z nanoparticles.

### 2.4 CPK-II: mass-energy transfer extension for local-dose and radiolysis enhancement

#### *2.4.1 $\bar{d}_1$: functional formulation*

The above-proposed CPK-I model starts from the most direct contributing factor, where the contribution of each individual nanoparticle to surface-mediated biochemical radiosensitization is assumed to depend directly on the mean exposed surface area per nanoparticle. From this assumption, an analytical expression for the equivalent dose per nanoparticle $\bar{d}_1$, which is proportional to the effective surface area $\bar{s}_1$, is established.

In high-Z nanoparticle radiosensitization, however, both local dose enhancement and radiolytic enhancement are initiated by secondary-electron-mediated energy-transfer processes. Therefore, the dominant factor governing each nanoparticle's direct contribution to these enhancement effects is no longer the exposed surface area $\bar{s}_1$, but the mean energy emission $\bar{\varepsilon}_1$ released from each nanoparticle. Accordingly, in the subsequent CPK-II model for high-Z nanoparticle radiosensitization, the equivalent dose per nanoparticle $\bar{d}_1$ is assumed to be proportional to the mean energy emission per nanoparticle $\bar{\varepsilon}_1$ via a coefficient $k_\mathrm{m}$ with the dimension of inverse mass, expressed as

$$\bar{d}_1 = k_\mathrm{m}\bar{\varepsilon}_1 \tag{26}$$

The energy emission per nanoparticle $\bar{\varepsilon}_1$ is defined as the total energy emission $\bar{E}$ released by the intracellular nanoparticle agglomerates divided by the mean number of nanoparticles per cell $\bar{\lambda}$, namely

$$\bar{\varepsilon}_1 = \bar{E}/\bar{\lambda} \tag{27}$$

The total energy emission $\bar{E}$ is given by the product of the total energy transfer $\bar{E}_\mathrm{k}$ and the total energy emission ratio $\beta_k$ of the intracellular nanopartciel ensemble

$$\bar{E} = \bar{E}_\mathrm{k}\beta_k \tag{28}$$

where $\bar{E}_\mathrm{k}$, defined as the total kinetic energy of secondary electrons (i.e., mass integral of kerma), scales linearly with the total mass of irradiated nanoparticles $\bar{\lambda}m_0$ and the incident radiation dose $D$ via a dimensionless coefficient $\psi$

$$\bar{E}_\mathrm{k} = \psi D\bar{\lambda}m_0 \tag{29}$$

while $\beta_k$ is defined as the ratio of the total kinetic energy carried by escaping electrons to the total energy transfer $\bar{E}_\mathrm{k}$ (mass integral of kerma) mediated by secondary electrons originating from the nanoparticles.

Combining equations (26)-(29), the equivalent dose per nanoparticle $\bar{d}_1$ is formulated as

$$\bar{d}_1 = k_\mathrm{m}\psi D m_0\beta_k = K_\mathrm{m} D m_0\beta_k \quad (K_\mathrm{m}{=}k_\mathrm{m}\psi) \tag{30}$$

Two induction coefficients are multiplied to form the lumped coefficient $K_\mathrm{m}=k_\mathrm{m}\psi$: $k_\mathrm{m}$ from equation (26) has an inverse mass dimension; $\psi$ from equation (29) is dimensionless. The combined parameter $K_\mathrm{m}$ retains the inverse mass dimension inherited from $k_\mathrm{m}$.

The total energy emission ratio $\beta_k$, which depends on incident photon energy and the agglomeration extent of intracellular nanoparticles, can be expressed in a functional form $\beta_k(\xi; E_\gamma)$ where $\xi$ is a generalized variable characterizing nanoparticle agglomeration extent and $E_\gamma$ denotes the fixed incident photon energy. The agglomeration dependence of $\beta_k$ involves many geometric factors such as bulk density, cluster size and topological structure. Consequently, explicit formulations for these dependencies are intractable. These geometric effects are therefore lumped within the parameterized form $\beta_k(\xi; E_\gamma)$, with $\xi$ characterizing the agglomeration extent and $E_\gamma$ the photon energy. An equivalent formulation for $\beta_k(\xi; E_\gamma)$ based on the single-sphere energy emission ratio is introduced in section 2.4.2.

### 2.4.2. $\beta_k(\xi; E_\gamma) \rightarrow \beta_k(R; E_\gamma)$: single-sphere equivalence and formulation of related quantities

For intracellular nanoparticles with a mean number $\bar{\lambda}$ per cell under a fixed agglomeration state, the energy emission ratio $\beta_k$ increases and saturates toward unity ($\beta_k \to 1$) as photon energy $E_\gamma$ rises. At a fixed $E_\gamma$, $\beta_k$ exhibits a negative dependence on the agglomeration extent $\xi$: increasing $\xi$ leads to a monotonic decrease in $\beta_k$, from the maximum value for fully isolated nanoparticles to the minimum for a fully packed single solid sphere.

The agglomeration effect on $\beta_k$ vanishes only at the two limiting agglomeration extremes: fully isolated nanoparticles and a fully packed single sphere. Under these two boundary conditions, $\beta_k$ depends exclusively on the size of either the isolated single nanoparticle or the fully packed single solid sphere. Accordingly, a system of fully isolated single spheres with equal total mass and different sphere size is introduced as a reference to characterize the energy emission ratio $\beta_k$ of randomly agglomerated intracellular nanoparticles, drawing on the property that $\beta_k$ depends only on sphere size at these two limiting extremes.

The energy-emission ratio of such a single-sphere system equals that of its constituent individual spheres and is denoted as $\beta_k(R; E_\gamma)$, which is fully governed by the sphere radius $R$ under a given $E_\gamma$. The boundary values of $\beta_k(R; E_\gamma)$ at the single-nanoparticle limit $\beta_k(R_0; E_\gamma)$ and the fully packed single-sphere limit $\beta_k(R_0\bar{\lambda}^{1/3}; E_\gamma)$ match exactly with $\beta_k(\xi; E_\gamma)$ of the nanoparticle system at fully isolated and fully packed agglomeration states, i.e. $\beta_k(\xi_0; E_\gamma)$ and $\beta_k(\xi_1; E_\gamma)$. For a fixed $E_\gamma$, both functions decrease from the shared maximum $\beta_k(\xi_0; E_\gamma) = \beta_k(R_0; E_\gamma)$ to shared minimum $\beta_k(R_0\bar{\lambda}^{1/3}; E_\gamma) = \beta_k(\xi_1; E_\gamma)$ nonlinearly with increasing agglomeration extent $\xi$ or single-sphere radius $R$. For a fixed agglomeration extent $\xi$ or radius $R$, both quantities rise from the shared minimum $\beta_k(R; 0) = \beta_k(\xi; 0) = 0$ at $E_\gamma = 0$ toward unity ($\beta_k \to 1$) with increasing $E_\gamma$. Although the $\xi$-dependence of $\beta_k(\xi; E_\gamma)$ may not be identical in functional form to the $R$-dependence of $\beta_k(R; E_\gamma)$, a one-to-one correspondence that preserves identical $\beta_k$ values can be established between each agglomeration extent and the corresponding single-sphere radius for a given $E_\gamma$.

From the discussion above, an equal-mass system of fully-isolated single spheres can be constructed to match the $\beta_k$ of the intracellular nanoparticles. For intracellular nanoparticles with a mean number $\bar{\lambda}$ per cell under fixed photon energy $E_\gamma$, the single-sphere radius can be formulated as $R_0\lambda^t$ with reference to the two limiting radii $R_0$ and $R_0\bar{\lambda}^{1/3}$, where the exponent $t \in (0, 1/3)$ serves as a phenomenological agglomeration-characterizing factor. Accordingly, for intracellular nanoparticles at agglomeration extent $\xi$, the $\beta_k$ can be equivalently described using that of the equal-$\beta_k$ single-sphere system

$$\beta_k(\xi; E_\gamma),\ \ \xi \in (\xi_0, \xi_1) \ \rightarrow \ \beta_k(R_0\bar{\lambda}^t; E_\gamma),\ t \in (0, 1/3) \tag{31}$$

Within this equivalence framework linking $\beta_k(\xi; E_\gamma)$ for nanoparticle systems and $\beta_k(R; E_\gamma)$ for fully-isolated equal-mass single sphere systems, analytical expressions for the interrelated quantities $\beta_k$, $\bar{E}$, $\bar{\varepsilon}_1$ and $\bar{d}_1$ are derived for the intracellular nanoparticles with a per-cell loading $\bar{\lambda}$. The expressions are given for fully isolated, fully packed, and arbitrary intermediate agglomeration states, as summarized in Table 1.

**Table1.** Analytical expressions for the correlated quantities $\beta_k, \bar{E}, \bar{\varepsilon}_1, \bar{d}_1$ of intracellular nanoparticles under distinct agglomeration states. $\beta_k$ denotes the total energy emission ratio, $\bar{E}$ the total energy emission of intracellular nanoparticles. $\bar{\varepsilon}_1$ the mean energy emission per nanoparticle, and $\bar{d}_1$ the equivalent dose per nanoparticle. $\bar{\lambda} = \bar{\lambda}(C)$ is the mean number of nanoparticles per cell. (a) Fully isolated state with minimum agglomeration ($t=0$). (b) Fully packed single solid sphere with maximum agglomeration ($t=1/3$). (c) Random dispersion with intermediate agglomeration, $t \in (0, 1/3)$.

(a)

| | |
|---|---|
| Total energy emission ratio | $\beta_k = \beta_k(R_0; E_\gamma)$ |
| Total energy emission | $\bar{E} = \bar{E}_k\beta_k = \psi D\bar{\lambda} m_0 \beta_k(R_0; E_\gamma)$ |
| Mean energy emission per nanoparticle | $\bar{\varepsilon}_1 = \bar{E}/\bar{\lambda} = \psi D m_0 \beta_k(R_0; E_\gamma)$ |
| Equivalent dose per nanoparticle | $\bar{d}_1 = k_m\bar{\varepsilon}_1 = k_m\psi D m_0 \beta_k(R_0; E_\gamma) = K_m D m_0 \beta_k(R_0; E_\gamma)$ |

(b)

| | |
|---|---|
| Total energy emission ratio | $\beta_k = \beta_k(R_0\bar{\lambda}^{1/3}; E_\gamma)$ |
| Total energy emission | $\bar{E} = \bar{E}_k\beta_k = \psi D\bar{\lambda} m_0 \beta_k(R_0\bar{\lambda}^{1/3}; E_\gamma)$ |
| Mean energy emission per nanoparticle | $\bar{\varepsilon}_1 = \bar{E}/\bar{\lambda} = \psi D m_0 \beta_k(R_0\bar{\lambda}^{1/3}; E_\gamma)$ |
| Equivalent dose per nanoparticle | $\bar{d}_1 = k_m\bar{\varepsilon}_1 = k_m\psi D m_0 \beta_k(R_0\bar{\lambda}^{1/3}; E_\gamma) = K_m D m_0 \beta_k(R_0\bar{\lambda}^{1/3}; E_\gamma)$ |

(c)

| | |
|---|---|
| Total energy emission ratio | $\beta_k = \beta_k(R_0\bar{\lambda}^t; E_\gamma) \quad t \in (0, 1/3)$ |
| Total energy emission | $\bar{E} = \bar{E}_k\beta_k = \psi D\bar{\lambda} m_0 \beta_k(R_0\bar{\lambda}^t; E_\gamma)$ |
| Mean energy emission per nanoparticle | $\bar{\varepsilon}_1 = \bar{E}/\bar{\lambda} = \psi D m_0 \beta_k(R_0\bar{\lambda}^t; E_\gamma)$ |
| Equivalent dose per nanoparticle | $\bar{d}_1 = k_m\bar{\varepsilon}_1 = k_m\psi D m_0 \beta_k(R_0\bar{\lambda}^t; E_\gamma) = K_m D m_0 \beta_k(R_0\bar{\lambda}^t; E_\gamma)$ |

*2.4.3. $\beta_k(R; E_\gamma)$: Radius-dependence calibration*

The radius-dependence of the single-sphere energy-emission ratio was calibrated via Monte Carlo simulation using TOPAS 4.1.p1 (Faddegon et al 2020, Perl et al 2012). A solid sphere of G4_Au material was modeled in vacuum, with radii ranging from 2.5 to 500 nm (specifically 2.5, 5, 10, 20, 30, 40, 50, 100, 200, 300, 400, and 500 nm). Separate beam configurations were implemented for a nominal 160 kVp source and three selected monoenergetic photon sources (5 keV, 500 keV, and 1 MeV). Each photon beam had an circular profile matched to the gold sphere diameter and zero-emittance characteristics, directed towards the gold sphere with a fixed source-to-center distance of 550 nm. $10^9$ incident photons were simulated for each energy condition.

The total energy transfer $\bar{E}_k$ (mass integral of kerma, equation (29)) mediated by the gold sphere was tallied using a custom volume scorer named"ScoreKineticEnergy", defined over the full sphere volume. A phase-space scorer was placed on the outer surface of the gold sphere to record escaping electrons and calculate their total kinetic energy. The TOPAS physics module g4em-penelope was used with Auger process enabled. Electromagnetic physics tables were generated for an energy range spanning 10 eV to 160 MeV, and a uniform production-cut of 1 nm was applied to all particles.

For the given $E_\gamma$, the simulated $\beta_k$ at different sphere radii were well described by an analytical radius-dependent function

$$\beta_k(R; E_\gamma) = a(R + d)^b \tag{32}$$

where $a$, $b$, and $d$ are all $E_\gamma$-dependent fitting parameters. The simulated data and fitted curves are provided in Appendix A.

*2.4.4. Relative corrections α"/α and β"/β: formulation, physical meaning and simplification*

Substituting the radius-dependent form $\beta_k(R_0\bar{\lambda}^t; E_\gamma)$ into equation (30) yields an explicit expression for the equivalent dose per nanoparticle at a given photon energy $E_\gamma$

$$\bar{d}_1(D, C) = \bar{d}_1\left(D, \bar{\lambda}(C)\right) = K_{\mathrm{m}} D m_0 \beta_k\left(R_0(\bar{\lambda}(C))^t; E_\gamma\right) \tag{33}$$

This expression shows that $\bar{d}_1$ depends linearly on both the single-nanoparticle mass $m_0$ and the incident photon dose $D$, and nonlinearly on the nanoparticle concentration $C$. The nonlinearity arises from the radius dependence of energy emission ratio $\beta_k(R; E_\gamma)$, in which concentration $C$ enters through $\bar{\lambda}(C)$ within the power-law argument of $\beta_k\left(R_0(\bar{\lambda}(C))^t; E_\gamma\right)$.

By substituting equation (33) (with $\bar{\lambda}(C)$ abbreviated as $\bar{\lambda}$) into (17) and subsequently into (15), the modified cell survival probability is obtained by assuming independent additive LQ killing (neglecting quadratic cross terms) and employing the equivalent doses of individual nanoparticles per cell, as derived below

$$\begin{aligned} S(D, C) &= e^{-(\alpha D + \beta D^2)} e^{-\bar{\lambda}\left[\alpha\left(K_{\mathrm{m}} D m_0 \beta_k(R_0\bar{\lambda}^t; E_\gamma)\right) + \beta\left(K_{\mathrm{m}} D m_0 \beta_k(R_0\bar{\lambda}^t; E_\gamma)\right)^2\right]} \\ &= e^{-(\alpha D + \beta D^2)} e^{-(\alpha'' D + \beta'' D^2)} = e^{-[(\alpha + \alpha'')D + (\beta + \beta'')D^2]} \end{aligned} \tag{34}$$

Similar to equation (21), the relative corrections $\alpha''/\alpha$ and $\beta''/\beta$ for the intrinsic LQ parameters $(\alpha, \beta)$ are derived as

$$\frac{\alpha''}{\alpha} = \bar{\lambda}\left(K_{\mathrm{m}} m_0 \beta_k(R_0\bar{\lambda}^t;\, E_\gamma)\right), \quad \frac{\beta''}{\beta} = \bar{\lambda}\left(K_{\mathrm{m}} m_0 \beta_k(R_0\bar{\lambda}^t;\, E_\gamma)\right)^2 \tag{35}$$

and simplied as

$$\frac{\alpha''}{\alpha} = \bar{\lambda}\left(K_{\mathrm{m}} m_0 \beta_k(R_0\bar{\lambda}^t;\, E_\gamma)\right), \quad \frac{\beta''}{\beta} \approx 0 \tag{36}$$

This simplification adopts the same physical assumption as used in equation (23): the induction coefficient of the equivalent dose per nanoparticle satisfies $\varphi_1 = \bar{d}_1/D \ll 1$ and $\alpha''/\alpha = \bar{\lambda}\varphi_1 < 1$, therefore $\beta''/\beta = (\alpha''/\alpha)\varphi_1 \ll \alpha''/\alpha < 1$.

By substituting equation (16) for $\bar{\lambda}(C)$ back into (36), the relative correction $\alpha''/\alpha$ is derived as a four-parameter function of the nanoparticle concentration $C$

$$\frac{\alpha''}{\alpha} = K_{\mathrm{m}} m_0 A(1 - e^{-kC})\beta_k\left[R_0(A(1 - e^{-kC}))^t;\, E_\gamma\right] \tag{37}$$

with free parameters $K_{\mathrm{m}} m_0 A$, $R_0 A^t$, $k$ and $t$. Parameter degeneracy for this expression, as well as that for the generalized dual-enhancement function, is addressed in section 2.4.5, with a summary provided in section 2.5.7.

*2.4.5 Generalized expression of dual-enhancement effects and its parameter degeneracy*

The radiosensitization effect of low-Z nanomaterials mainly originates from surface-mediated enhancement of biochemical effects, including intrinsic cellular oxidative stress induced at nanomaterial surfaces, such as endogenous reactive oxygen species accumulation, as well as enhanced biological responses from functionalized surfaces. For the radiosensitization by high-Z nanoparticles, strictly speaking, the enhancement effects arise not only from the local dose and radiolytic free radical production via energy transfer of secondary electrons, but also from surface-mediated biochemical enhancement which can be inherently induced by most types of nanomaterials. Therefore, it is necessary to construct a generalized model to describe LQ parameter corrections accounting for the dual ehancement effects of high-Z nanoparticles.

Combining equation (19) of the CPK-I model for surface-mediated biochemical enhancement and equation (33) of the CPK-II model for mass-energy-transfer-mediated enhancement yields a general expression for the equivalent dose per nanoparticle describing the dual enhancement at a fixed photon energy $E_\gamma$

$$\bar{d}_1 = \bar{d}_{1s} + \bar{d}_{1m} = K_s D s_0 \bar{\lambda}^{\eta-1} + K_m D m_0 \beta_k\left(R_0 \bar{\lambda}^t;\ E_\gamma\right) \tag{38}$$

By substituting equation (38) into (17) and the resulting expression into (15), the LQ parameter modification arising from the dual enhancement effects is derived as a six-parameter function of nanoparticle concentration $C$

$$\frac{\alpha''}{\alpha} = K_s s_0 \bar{\lambda}^\eta + \bar{\lambda} K_m m_0 \beta_k\left(R_0 \bar{\lambda}^t;\ E_\gamma\right) = K_1\left(1-e^{-kC}\right)^\eta + K_2\left(1-e^{-kC}\right)\beta_k\left[\theta\left(1-e^{-kC}\right)^t;\ E_\gamma\right] \tag{39}$$
$$\left(K_1 = K_s s_0 A^\eta,\ \ K_2 = K_m m_0 A,\ \ \theta = R_0 A^t\right)$$

in which the six fitting parameters are $K_1 = K_s s_0 A^\eta$, $K_2 = K_m m_0 A$, $\theta = R_0 A^t$, $k$, $\eta$ and $t$. The lumped coefficients $K_s s_0 A^\eta$ and $K_m m_0 A$ for the two enhancement mechanisms, as well as $\theta = R_0 A^t$, each comprise multiple physical constants and can only be calibrated jointly. Meanwhile, $k$, $\eta$ and $t$ are kinetic and nonlinear exponent parameters that can be fitted independently. It is clear that this six-parameter function of the nanoparticle concentration $C$ is impractical for predicting the LQ parameter modification associated with the dual enhancement effects.

In practice, the primary particle size $R_0$ and the saturation coefficient $A$ for the mean number of nanoparticles per cell $\bar{\lambda}(C)$ are experimentally quantifiable. Once $R_0$ and $A$ are determined, $\theta = R_0 A^t$ is constrained by the two fixed constants and the exponent $t$, thus no longer serving as a free fitting parameter. Meamwhile, the primary single-nanoparticle surface area $s_0$ and mass $m_0$ can be uniquely calculated from $R_0$ and are thus also treated as known constants. As a result, the model's free fitting parameters are reduced from six to five, namely $K_s$, $K_m$, $k$, $\eta$ and $t$.

Further reduction of model parameters can only be achieved by extra constraints on the five parameters to cut fitting degrees of freedom, via two pathways: (1) A mechanistic relation $K_m = \gamma K_s$ with pre-determined coefficient $\gamma$ unifies the amplitude coefficients for the two enhancement effects into a single term, lowering the parameter count from five to four; (2) Deriving an inherent relationship linking $t$ and $\eta$ reduces the two exponential coefficients, yielding a four-parameter model.

The main difficulty for the first approach $K_m = \gamma K_s$ lies in the following aspects. First, $K_s$ and $K_m$ describe two independent biological and physical mechanisms, namely surface-mediated radiosensitization and mass-energy transfer mediated local dose and radiolysis enhancement, there is no universal intrinsic proportionality constant $\gamma$. Second, the calibration requires elaborate decoupling control experiments, yet the obtained $\gamma$ value is only valid for specific conditions, lacking universality. In contrast, the analytical relation between $t$ and $\eta$, stemming from unified nanoparticle uptake-agglomeration kinetics, is simpler to derive without artificial biased assumptions.

## 2.5 Physical t-η correlation and parameter-reduced predictive models

### *2.5.1 Two equivalent single-sphere systems: equal energy-emission ratio and equal total effective area*

In a general sense, under the constraint of idential total mass, a one-to-one mapping can always be established between an agglomerated nanoparticle system (with agglomeration bounded between fully isolated and fully packed limits) and a fully isolated single-sphere system (with the sphere size ranging from the primary nanoparticle size up to the size of fully packed single solid sphere). Different equivalence-matching criteria yield distinct mappings between these two configurations.

Under the equal-$\beta_k$criterion at fixed $E_\gamma$, the agglomeration-dependent $\beta_k(\xi; E_\gamma)$ of a $\bar{\lambda}$-nanoparticle system and the radius-dependent $\beta_k(R; E_\gamma)$ of the equal-mass single sphere system constitute a pointwise bijective mapping with boundary values $\beta_k(\xi_0; E_\gamma) = \beta_k(R_0; E_\gamma)$ and $\beta_k(\xi_1; E_\gamma) = \beta_k(R_0 \bar{\lambda}^{1/3}; E_\gamma)$. Substituting $\beta_k(\xi; E_\gamma)$ into the inverse function of $\beta_k(R; E_\gamma)$, i.e., $\beta_k^{-1}(\beta_k; E_\gamma) = R_\beta(\beta_k; E_\gamma)$, yields the equal-$\beta_k$ single-sphere radius $R_\beta(\beta_k(\xi; E_\gamma); E_\gamma)$which carries implicit $E_\gamma$ dependence. Substituting it back into $\beta_k(R; E_\gamma)$ gives the mapping identity $\beta_k(\xi; E_\gamma) = \beta_k[R_\beta(\beta_k(\xi; E_\gamma); E_\gamma); E_\gamma]$.
In parallel, under the equal-$\bar{S}$ criterion, the agglomerationdependent $\bar{S}(\xi)$ for $\bar{\lambda}$-nanoparticle systems and radius-dependent $\bar{S}(R)$ of the fully isolated equal-mass single-sphere systems — whose explicit formula is derived as

$$\bar{S}(R) = \frac{4\pi R^2 \bar{\lambda}}{(4\pi R^3/3)/(4\pi R_0^3/3)} = 4\pi R_0^2 \bar{\lambda}\frac{R_0}{R} = s_0 \bar{\lambda}\frac{R_0}{R}, \quad R \in \left(R_0, R_0 \bar{\lambda}^{1/3}\right) \tag{40}$$

also constitute a pointwise bijective mapping with identical boundary values $\bar{S}(\xi_0) = \bar{S}(R_0) = s_0 \bar{\lambda}$, $\bar{S}(\xi_1) = \bar{S}(R_0 \bar{\lambda}^{1/3}) = s_0 \bar{\lambda}^{2/3}$. Substituting $\bar{S}(\xi)$ into the inverse function of $\bar{S}(R)$, i.e., $\bar{S}^{-1}(\bar{S}) = R_S(\bar{S}) = s_0 \bar{\lambda} R_0/\bar{S}$, yields the equal-$\bar{S}$ single-sphere radius $R_S(\bar{S}(\xi)) = s_0 \bar{\lambda} R_0/\bar{S}(\xi)$. Substituting it back into $\bar{S}(R)$ gives the mapping identity $\bar{S}(\xi) = S(R_S(\bar{S}(\xi))) = \bar{S}(s_0 \bar{\lambda} R_0/\bar{S}(\xi))$.

### 2.5.2 Two equivalent single-sphere radii: the $R_\beta$-$R_S$ correlation governing the t–η relationship

As outlined above, for a $\bar{\lambda}$-nanoparticle system at an agglomeration extent $\xi$ and a fixed photn energy $E_\gamma$, the equal-$\beta_k$ and equal-$\bar{S}$ radii of a fully isolated equal-mass single-sphere system are generally expressed as

$$R_\beta = R_\beta(\beta_k(\xi; E_\gamma); E_\gamma), \quad R_S = R_S(\bar{S}(\xi)) = s_0\bar{\lambda}\, R_0/\bar{S}(\xi) \tag{41}$$

In section 2.3.1, the total effective area of a $\bar{\lambda}$-nanoparticle system is formulated as $\bar{S}=s_0\bar{\lambda}^{\eta}, \eta \in (2/3, 1)$, with the boundary values $\bar{S}=s_0\bar{\lambda}$ for fully isolated nanoparticles, and $\bar{S}=s_0\bar{\lambda}^{2/3}$ for the fully packed single sphere. In section 2.4.2, the energy emission ratio at a fixed $E_\gamma$ is given as $\beta_k=\beta_k(R_0\bar{\lambda}^t; E_\gamma), t \in (0,1/3)$, referring to the boundary values $\beta_k(R_0; E_\gamma)$ for fully isolated nanoparticles and $\beta_k(R_0\bar{\lambda}^{1/3}; E_\gamma)$ for the fully packed single sphere.

Substituting $\beta_k=\beta_k(R_0\bar{\lambda}^t; E_\gamma)$ into $R_\beta=R_\beta(\beta_k(\xi; E_\gamma); E_\gamma)$ from equation (41) yields the $t$-dependent expression for the equal-$\beta_k$ single-sphere radius

$$R_\beta = R_\beta(\beta_k(R_0\bar{\lambda}^t; E_\gamma); E_\gamma) = R_0\bar{\lambda}^t, \quad t \in (0,1/3) \tag{42}$$

Note that $R_\beta=R_0\bar{\lambda}^t$ does not imply independence from $E_\gamma$, since its $E_\gamma$-dependence enters via the parameter $t$.

Substituting $\bar{S}=s_0\bar{\lambda}^{\eta}$ into $R_S=R_S(\bar{S}(\xi))=s_0\bar{\lambda}\, R_0/\bar{S}(\xi)$ from equation (41) yields the $\eta$-dependent expression for the equal-$\bar{S}$ single-sphere radius

$$R_S = R_S(s_0\lambda^{\eta}) = s_0\bar{\lambda}\, R_0/s_0\bar{\lambda}^{\eta} = R_0\bar{\lambda}^{1-\eta}, \quad \eta \in (2/3, 1) \tag{43}$$

From equations (42) and (43), the physical relation between the phenomenological parameters $t$ and $\eta$—separately defined for characterizing the agglomeration extent in CPK-II and the dispersion extent in CPK-I—is intrinsically governed by the correlation between the equivalent single-sphere radii $R_\beta$ and $R_S$ for the equal-$\beta_k$ and equal-$\bar{S}$ single-sphere systems. This directly determines the inherent $t$-$(1-\eta)$ relationship via the derived expressions $R_\beta=R_0\bar{\lambda}^t$ and $R_S=R_0\bar{\lambda}^{1-\eta}$.

### 2.5.3 Intrinsic inequality $R_\beta > R_S$ and the resultant $t$-$(1-\eta)$ relationship

To reveal the intrinsic correlation between the single-sphere radii $R_\beta$ and $R_S$, Monte Carlo simulations were performed on randomly agglomerated gold nanoparticle systems with 30 nm and 500 nm diameters across different agglomeration extents. The simulations considered 100-nanoparticle agglomerates in the fully isolated and fully packed limits, together with three intermediate agglomeration configurations: Group 1 consisted of one random agglomerate containing 100 nanoparticles; Group 2 contained two identical 50-nanoparticle clusters simulated separately, representing two fully isolated agglomerates with infinite spatial separation; Group 3 contained five identical 20-nanoparticle clusters that were also simulated separately, representing five fully isolated 20-nanoparticle agglomerates. The agglomeration extent of the five configurations followed the ascending order: fully isolated nanoparticles < Group 3 < Group 2 < Group 1 < fully packed solid sphere.

The effective area $\bar{S}$, and the energy-emission ratio $\beta_k$ for different photon energies of 5 keV, 160 kVp, 500 keV and 1 MeV, were computed for these nanoparticle systems. Substituting the calculated $\bar{S}$ and $\beta_k$ data into equation (41) yielded the $R_S$ value and the $E_\gamma$-dependent $R_\beta$ values, from which the $R_\beta$-$R_S$ correlation for the five agglomeration extents, each evaluated at four photon energies, was obtained for the 30 nm and 500 nm GNP agglomerate models. The resulting $R_\beta$-$R_S$ correlation data points are shown in figure 1. The generation procedure for the agglomerated GNPs models and the calculation methods for the total effective surface area $\bar{S}$ and energy emission ratio $\beta_k$ of the nanoparticle systems are described in Appendix B.

The results in figure 1 indicate that, for both the 30 nm and 500 nm gold nanoparticles across the five agglomeration extents and four photon energies, the equal-$\beta_k$ and equal-$\bar{S}$ single-sphere radii satisfy $R_\beta=R_\beta(\beta_k(\xi; E_\gamma); E_\gamma)>R_S=R_S(\bar{S}(\xi))\geq R_0$, or simply, $R_\beta>R_S\geq R_0$. From this intrinsic inequality, it can be deduced that the two phenomenological parameters $t$ and $\eta$ consistently satisfy $t>1-\eta$ within their shared value range (0, 1/3). The fully isolated state gives $t=(1-\eta)=0$, while the fully packed sphere corresponds to $t=(1-\eta)=1/3$. Substituting the simulated $R_\beta$ and $R_S$ values into equations (42) and (43) yields the paired values of $t$ and $1-\eta$. For the five agglomeration configurations, the $1-\eta$ values derived from the $R_S$ data and their corresponding $t$ values from the $R_\beta$ data evaluated at the monoenergetic photon energies of 5 keV, 500 keV and 1 MeV are summarized in table 2 and plotted in figure 2a and 2b.

The fitted curves characterizing the $t$-$(1-\eta)$ correlations are plotted in figure 2c and 2d. All curve-fitting parameters are summarized in Table 3. As shown in the figure, the $t$-$(1-\eta)$ relationship corrersponds well to the equivalent single-sphere radius inequality $R_\beta>R_S\geq R_0$. The fitted $t$-$(1-\eta)$ curves and underlying $R_\beta$-$R_S$ curves (not shown) are both convex, with identical values at both boundaries and $t>1-\eta$ ($R_\beta>R_S$) for all intermediate states.

Partial variation trends of the $R_\beta$-$R_S$ correlation and corresponding $t$-$(1-\eta)$ relationships with varying agglomeration extent and photon energy can also be seen in figures 1 and 2. A full analysis of the photon-energy and agglomeration-extent dependences of the $R_\beta$-$R_S$ relationship and their disparity is presented in the following section 2.5.4.

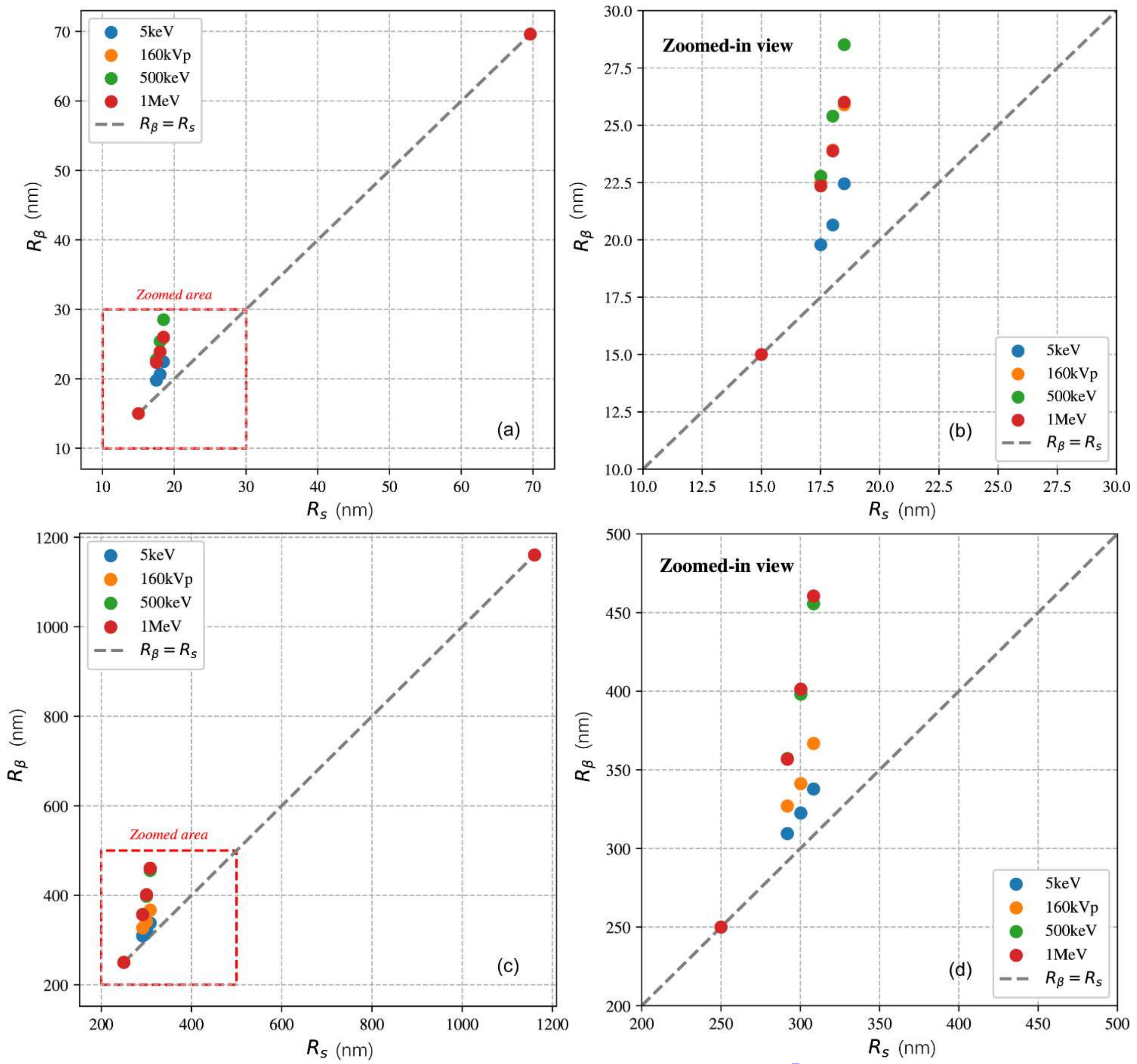

**Figure 1**. Correlated data points showing the $R_\beta$-$R_S$ relation between the equal-$\beta_k$ and equal-$\bar{S}$ single-sphere radii $R_\beta$ and $R_S$, obtained for the 30 nm (a,b) and 500 nm (c,d) gold nanoparticle systems across five agglomeration extents under four photon energies (5 keV, 500 keV, 1 MeV, and 160 kVp). The dashed line denotes $R_\beta$=$R_S$. The right panels show zoomed-in views of the regions marked by the red dashed boxes in the left panels.

**Table 2.** Phenomenological parameters of $t$ and $1-\eta$ derived from the $R_{\beta}$ and $R_{S}$ data (given in figure 1) via $R_{\beta}=R_0\bar{\lambda}^t$ and $R_S=R_0\bar{\lambda}^{1-\eta}$. The correrlated data points are presented for gold nanoparticles with primary diameters of 30 nm and 500 nm, across five agglomeration states and three monoenergetic photon energies.

| $1-\eta$ | $t$ ($D$ =30nm, $E_{\gamma}$= 5keV) | $t$ ($D$ =30nm, $E_{\gamma}$= 500keV) | $t$ ($D$ =30nm, $E_{\gamma}$= 1MeV) |
|---|---|---|---|
| 0 | 0 | 0 | 0 |
| 0.033605 | 0.060171969 | 0.090703409 | 0.086621435 |
| 0.039803 | 0.069400825 | 0.114359274 | 0.100984415 |
| 0.045556 | 0.087555867 | 0.139488382 | 0.119476704 |
| 0.333333 | 0.333333333 | 0.333333333 | 0.333333333 |
| $1-\eta$ | $t$ ($D$ =500nm, $E_{\gamma}$= 5keV) | $t$ ($D$ =500nm, $E_{\gamma}$= 500keV) | $t$ ($D$ =500nm, $E_{\gamma}$= 1MeV) |
| 0 | 0 | 0 | 0 |
| 0.033605 | 0.046315697 | 0.077461815 | 0.077204408 |
| 0.039803 | 0.055309951 | 0.101012498 | 0.102743579 |
| 0.045556 | 0.065372224 | 0.130309609 | 0.132619817 |
| 0.333333 | 0.333333333 | 0.333333333 | 0.333333333 |

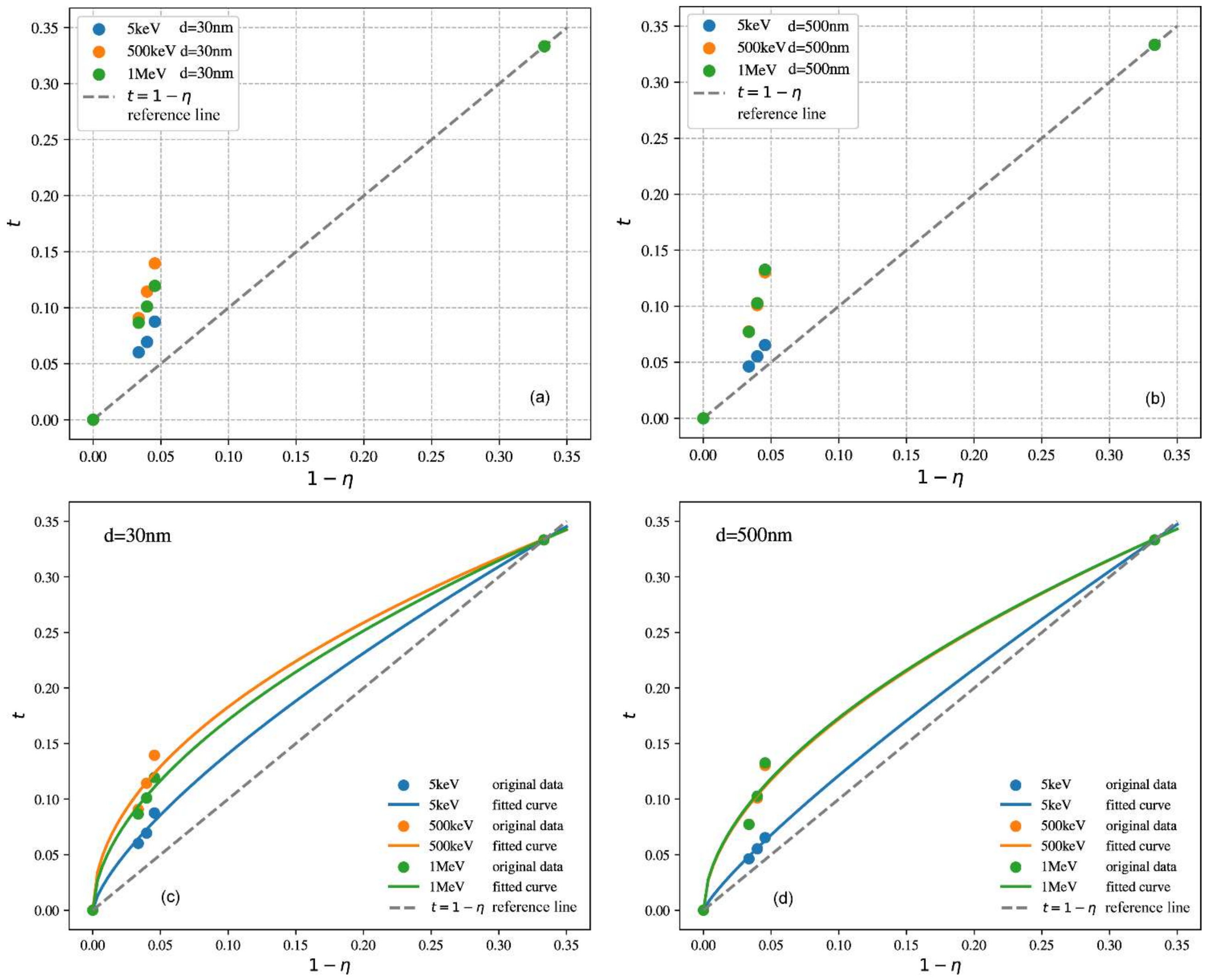

**Figure 2**. Correlated data points and fitted curves illustrating the $t$-$(1-\eta)$ parameter relationship for the 30 nm (a,c) and 500 nm (b,d) gold nanoparticles. All datasets cover five agglomeration configurations and three photon energies. Panels (a,b) show raw data points, while panels (c,d) show the corresponding fitted curves.

**Table 3.** Fitting parameters for the $t$-$(1-\eta)$ correlation curves plotted in figure 2b and 2d for the 30 nm and 500 nm gold nanoparticles across five agglomeration configurations and three monoenergetic photon energies.

| Fitted fuction | $D$ =30nm, $E_\gamma$= 5keV | $D$ =30nm, $E_\gamma$= 500keV | $D$ =30nm, $E_\gamma$= 1MeV |
|---|---|---|---|
| $t = a(1-\eta)^b$ | a = 0.732370<br>b = 0.715999<br>$R^2$ = 0.998358<br>Adjusted $R^2$ = 0.995075 | a = 0.578963<br>b = 0.500457<br>$R^2$ = 0.985296<br>Adjusted $R^2$ = 0.955888 | a = 0.613164<br>b = 0.553855<br>$R^2$ = 0.996852<br>Adjusted $R^2$ = 0.990556 |
| | $D$ =500nm, $E_\gamma$= 5keV | $D$ =500nm, $E_\gamma$= 500keV | $D$ =500nm, $E_\gamma$= 1MeV |
| | a = 0.838510<br>b = 0.839522<br>$R^2$ = 0.999782<br>Adjusted $R^2$ = 0.999346 | a = 0.612381<br>b = 0.551500<br>$R^2$ = 0.981943<br>Adjusted $R^2$ = 0.945830 | a = 0.608727<br>b = 0.545874<br>$R^2$ = 0.982061<br>Adjusted $R^2$ = 0.946183 |

*2.5.4 Direct cause for $R_\beta > R_S$: energy-emission ratio disparity between nanoparticle agglomerates and equal-$\bar{S}$ single-sphere systems*

The direct cause of the single-sphere radius inequality $R_\beta > R_S$ lies in the intrinsic disparity in energy emission ratio between two systems: for any agglomeration extent $\xi$ and photon energy $E_\gamma$, nanoparticle systems always have a lower $\beta_k$ than their equal-$\bar{S}$ single-sphere counterparts at the same $E_\gamma$, written as

$$\beta_k(\xi; E_\gamma) < \beta_k[R_S(\bar{S}(\xi)); E_\gamma] \tag{44}$$

This inequality directly gives rise to the radius inequality $R_\beta > R_S$ and the corresponding parameter inequality $t > 1 - \eta$ for the equal-$\beta_k$ and equal-$\bar{S}$ single-sphere systems.

This statement can be explained using the full expression of equation (44)

$$\beta_k(\xi; E_\gamma) = \beta_k\left[R_\beta\left(\beta_k(\xi; E_\gamma); E_\gamma\right); E_\gamma\right] < \beta_k\left[R_S\left(\bar{S}(\xi)\right); E_\gamma\right] \tag{45}$$

which states that the $\beta_k$ of the equal-$\beta_k$ single-sphere systems is invariably lower than that of equal-$\bar{S}$ single-sphere systems at any $E_\gamma$. Since the single-sphere energy emission ratio function $\beta_k(R; E_\gamma)$ decreases monotonically with $R$, equation (45) implies that at any $E_\gamma$, the radii of the equal-$\beta_k$ and equal-$\bar{S}$ single-sphere systems always satisfy

$$R_\beta(\beta_k(\xi; E_\gamma); E_\gamma) > R_S\left(\bar{S}(\xi)\right) \tag{46}$$

and thus yield the parameter inequality $t > 1 - \eta$. Therefore, it can be stated that the invariably lower energy emission ratio of nanoparticle systems relative to that of their equal-$\bar{S}$ single-sphere systems (equation (44)) is the direct cause for $R_\beta > R_S$.

For subsequent analysis, the energy emission ratio of a nanoparticle system at agglomeration extent $\xi$, and that of its equal-$\bar{S}$ single-sphere counterpart at the same photon energy $E_\gamma$, are respectively abbreviated as

$$\beta_\xi = \beta_k\left(\xi; E_\gamma\right); \quad \beta_S = \beta_k[R_S\left(\bar{S}(\xi)\right); E_\gamma] \tag{47}$$

The disparity in energy-emission ratio, and the resultant disparity in single-sphere radius, are respectively written as

$$\Delta\beta_k = \beta_S - \beta_\xi = \beta_k\left[R_S\left(\bar{S}(\xi)\right); E_\gamma\right] - \beta_k\left(\xi; E_\gamma\right); \quad \Delta R = R_\beta - R_S = R_\beta(\beta_k\left(\xi; E_\gamma\right); E_\gamma) - R_S\left(\bar{S}(\xi)\right) \tag{48}$$

The following discussion addresses how $\Delta\beta_k$ and the resultant $\Delta R$ vary with photon energy and nanoparticle agglomeration. A qualitative reasoning on the $E_\gamma$-dependence of $\Delta\beta_k$ can be performed first. For a nanoparticle system at agglomeration extent $\xi$ and its equal-$\bar{S}$ single-sphere system with radius $R_S$, the energy-emission ratios $\beta_\xi$ and $\beta_S$ both equal zero at $E_\gamma = 0$, where $\Delta\beta_k = 0$, and both rise nonlinearly toward unity $(\beta_k \to 1)$ with the increase of $E_\gamma$ and converge to the same saturation value $\beta_k \to 1$ as $E_\gamma \to \infty$, where $\Delta\beta_k \to 0$. Under the constraint of equation (44), we deduce that $\Delta\beta_k$ increases with rising $E_\gamma$, reaches a maximum at a specific $E_\gamma$, then declines and converges to zero as $E_\gamma \to \infty$. Figure 3a illustrates the $E_\gamma$-dependent trends of $\beta_\xi$ and $\beta_S$, as well as their disparity $\Delta\beta_k$, for a fixed agglomeration extent $\xi$, using schematic curves (not drawn to scale). From the $E_\gamma$-dependent trend of $\Delta\beta_k$ and the resulting constraint imposed by equation (46), the $E_\gamma$-dependent variation of $\Delta R$ can be predicted: $\Delta R \to 0$ for both boundary conditions $E_\gamma \to 0$ and $E_\gamma \to \infty$, and $\Delta R > 0$ for intermediate $E_\gamma$ value. In other words, $\Delta R$ also rises from zero to a peak value, then decreases and converges to zero as $E_\gamma$ increases.

The above-deduced variation of $\Delta\beta_k$ and the resultant trend of $\Delta R$ with $E_\gamma$ can be illustrated more intuitively. By mapping the $E_\gamma$-dependent curves of $\beta_\xi$ and $\beta_S$ (figure 3a) onto the $R$-dependent $\beta_k(R; E_\gamma)$ curves for varying $E_\gamma$ (figure 3b), the $\beta_S$ curve consistently corresponds to a constant radius $R_S = R_S\left(S(\xi)\right)$ of the equal-$\bar{S}$ single-sphere system, while the $\beta_\xi$ curve, lying below the $\beta_S$ curve, yields an $E_\gamma$-dependent radius $R_\beta = R_\beta\left(\beta_k\left(\xi; E_\gamma\right); E_\gamma\right)$ of the equal-$\beta_k$ single-sphere system that exceeds the constant $R_S$ and varies with $E_\gamma$, leading to the disparity $\Delta R = R_\beta - R_S$, which rises from zero at $E_\gamma = 0$, reaches a maximum at a specific $E_\gamma$, then converges to zero as $E_\gamma \to \infty$.

These $E_\gamma$-dependent evolutionary features of the energy-emission ratios $\beta_\xi$ and $\beta_S$, along with their disparity $\Delta\beta_k$, and the coupled evolution of equal-$\beta_k$ and equal-$\bar{S}$ single-sphere radii $R_\beta$, $R_S$, and their disparity $\Delta R = R_\beta - R_S$, can be generalized to the full range of agglomeration extent $\xi \in (\xi_0, \xi_1)$, from fully isolated state to fully packed single-solid sphere. As shown in figure 3c, the paired $\beta_k$-$E_\gamma$ curves share $\beta_k = 0$ at $E_\gamma = 0$ and a limiting value $\beta_k \to 1$ as $E_\gamma \to \infty$. They fully overlap along the upper boundary $\beta_k\left(\xi_0; E_\gamma\right) = \beta_k[R_S\left(\bar{S}(\xi_0)\right); E_\gamma]$ at the fully isolated limit $\xi = \xi_0$. As $\xi$ increases, the paired curves both decline overall and gradually diverge, yielding larger $\Delta\beta_k$ across all photon energies. When $\xi$ reaches the upper limit $\xi = \xi_1$ correspondig to the fully packed single-solid sphere, both $\beta_k$-$E_\gamma$ curves fall onto the lower boundary and coincide again at $\beta_k\left(\xi_1; E_\gamma\right) = \beta_k[R_S\left(\bar{S}(\xi_1)\right); E_\gamma]$, where $\Delta\beta_k$ vanishes entirely.

From figure 3c, it can be inferred that as the agglomeration extent rises from $\xi_0$ to $\xi_1$, the equal-$\bar{S}$ single-sphere radius $R_S$, obtained by mapping the $\beta_S$ curve to the radius-dependent $\beta_k(R;E_\gamma)$ curves at various $E_\gamma$, increases monotonically from $R_0$ to $R_0\bar{\lambda}^{1/3}$. The corresponding disparity $\Delta R = R_\beta - R_S$ also varies with increasing agglomeration extent over the entire $E_\gamma$ range: $\Delta R = 0$ at $\xi = \xi_0$, increases to a maximum, and finally decreases back to $\Delta R = 0$ at $\xi = \xi_1$.

Figure 3d shows the $\xi$-dependent curves of the $\beta_\xi$ and $\beta_S$ for fixed $E_\gamma$. As $\xi$ increases from $\xi_0$ to $\xi_1$, the paired $\beta_k$-$\xi$ curves decline from the coincident $\beta_k(\xi_0;E_\gamma) = \beta_k[R_S(\bar{S}(\xi_0));E_\gamma]$ to $\beta_k(\xi_1;E_\gamma) = \beta_k[R_S(\bar{S}(\xi_1));E_\gamma]$. Across the whole range of $\xi$, the $\beta_\xi$ curve always sits below $\beta_S$. The $\Delta\beta_k$ starts at 0 when $\xi = \xi_0$, rises to maximum at intermediate agglomeration, and finally converges to zero at $\xi = \xi_1$. As $E_\gamma$ rises, the $\xi$-dependent curves shift upward and exhibit weaker attenuation. This is consistent with the observed features of higher $\beta_k$ values, smaller disparity $\Delta\beta_k$, and a narrower range of $\beta_k$ at elevated $E_\gamma$, as illustrated in figure 3c.

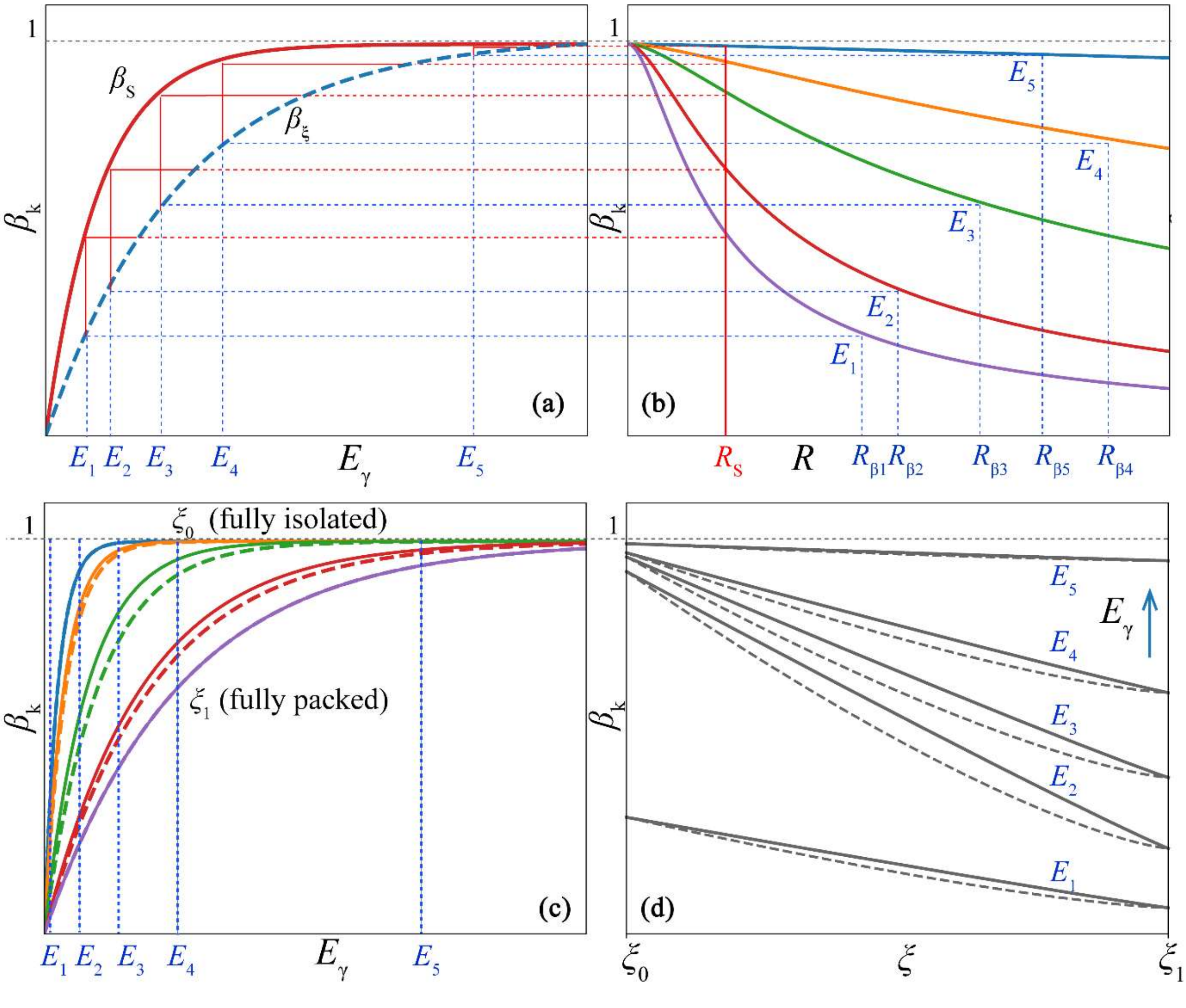

**Figure 3.** Schematic curves (not drawn to scale) illustrating the coupled variations of $\Delta\beta_k$— the energy emission ratio disparity between nanoparticle systems and their equal-$\bar{S}$ single-sphere systems—and the resultant $\Delta R$, the radius disparity between the equal-$\beta_k$ and equal-$\bar{S}$ single-sphere systems. (a) $E_\gamma$-dependent $\beta_\xi$ of a nanoparticle system at fixed $\xi$ and the $\beta_S$ of its equal-$\bar{S}$ single-sphere system, showing inequality $\beta_\xi < \beta_S$ ; (b) Radius-dependent single-sphere energy-emission ratio $\beta_k(R;E_\gamma)$ at different $E_\gamma$. Mapping the $\beta_k$-$E_\gamma$ curves from (a) onto the $\beta_k$-$R$ curves in (b) yields a constant $R_S$ and $E_\gamma$-dependent $R_\beta$, showing $R_\beta > R_S$; (c) $E_\gamma$-dependent $\beta_\xi$ and $\beta_S$ for nanoparticle systems across the agglomeration extent range $\xi \in (\xi_0, \xi_1)$; (d) $\xi$-dependent $\beta_\xi$ and $\beta_S$ curves at different fixed $E_\gamma$.

*2.5.5 Computational verification of the energy-emission ratio disparity and its evolution trends*

The verification used calculated data of the total effective area $\bar{S}$ and energy-emission ratio $\beta_k$ obtained in section 2.5.3 for 30 nm and 500 nm gold nanoparticle systems across the five agglomeration states and four photon energies. The $\bar{S}$ values are substituted into $R_\mathrm{S}(S(\xi))=s_0\bar{\lambda}\,R_0/\bar{S}(\xi)$ (equation (41)) to calculate the equal-$\bar{S}$ single-sphere radius $R_\mathrm{S}$. The $R_\mathrm{S}$ values are then substituted into the calibrated single-sphere energy-emission ratio function $\beta_k(R;E_\gamma)$ (equation (32)) to determine $\beta_S$ for the equal-$\bar{S}$ single-sphere system. A comparison of $\beta_S$ and the intrinsic $\beta_\xi$ is given in figure 4a, while the ratio $\beta_S/\beta_\xi$ is shown in figure 4b, covering the investigated agglomeration extents and photon energies. Figure 4c further illustrates the comparison between $\beta_\xi$ and $\beta_S$ by presenting their fitted $E_\gamma$-dependent curves at different agglomeration extents for the 500 nm nanoparticle system.

For each agglomeration extent, the energy-emission ratios $\beta_\xi$ and $\beta_S$ both increase with rising $E_\gamma$ and saturate toward unity. Notably, $\beta_\xi<\beta_S$ holds under all conditions (figure 4a). Accordingly, $\beta_S/\beta_\xi$ is greater than 1 for each agglomeration state, decreases as $E_\gamma$ rises, and gradually converges to unity (figure 4b). This trend is consistent with the features illustrated in figures 3a and 3c, governs the $E_\gamma$-dependent variation of $\Delta\beta_\mathrm{k}$ and further explains the evolution of the $R_\beta$-$R_\mathrm{S}$ relationship in figure 1: $R_\beta>R_\mathrm{S}$ holds for all agglomeration states; meanwhile, $\Delta R$ gradually saturates with increasing $E_\gamma$ in the energy range from 5 keV to 1 MeV. (Note that the $R_\beta$-$R_\mathrm{S}$ data sets shown in figure 1 do not cover higher photon energies. At higher energies, $\Delta R$ declines again and eventually approaches zero). The curves in figure 4c show that the $\beta_\mathrm{k}$ values at different agglomeration extents all rise rapidly at low photon energies and gradually plateau at higher energies, with clear disparities emerging between the nanoparticle systems and their equal-$\bar{S}$ single-sphere systems within the energy range 100–200 keV, as highlighted in the magnified inset. The curves in figure 4c fully validate the trends predicted in figure 3c.

At fixed $E_\gamma$, $\beta_\xi$ and $\beta_S$ both decrease monotonically as the agglomeration extent increases. For all cases, $\beta_\xi<\beta_S$ (figure 4a). The ratio $\beta_S/\beta_\xi$ first rises and then falls, starting and ending near unity. A higher $E_\gamma$ yields greater $\beta_k$ values but a smaller $\beta_S/\beta_\xi$ ratio (figure 4b). The disparity $\Delta\beta_\mathrm{k}$ initially increases and then decreases with increasing $\xi$, approaching zero at both the lower and upper limits. This trend matches the $\xi$-dependent variation of $\Delta\beta_\mathrm{k}$ deduced from figures 3c and 3d and explains the trend of the $R_\beta$-$R_\mathrm{S}$ relationship in figure 1: at fixed $E_\gamma$, $\Delta R$ first expands and then narrows as the agglomeration extent increases. At extremely high agglomeration approaching a solid single sphere, $\Delta R$ drops again and converge toward zero (not shown).

Based on the shcematic analysis in section 2.5.4 and computational verification in section 2.5.5, we reach a core conclusion: for nanoparticle systems in an arbitrary agglomeration state, the photon-irradiation-induced secondary-electrons energy emission ratio is always lower than that of fully-isolated, equal-mass and equal-area single-sphere systems under identical irradiation conditions, i.e., $\beta_\xi=\beta_k(\xi;E_\gamma)<\beta_S=\beta_k[R_\mathrm{S}(\bar{S}(\xi));E_\gamma]$. This intrinsic disparity constitutes the direct cause for the inequality between the equal-$\beta_k$ and equal-$\bar{S}$ single-sphere radii of nanoparticle systems $R_\beta>R_\mathrm{S}$ and the corresponding parameter inequality $t>1-\eta$. The evolution trends of the energy emission ratios $\beta_\xi=\beta_k(\xi;E_\gamma)$ and $\beta_S=\beta_k[R_S(\bar{S}(\xi));E_\gamma]$ and their disparity $\Delta\beta_\mathrm{k}=\beta_S-\beta_\xi$ govern the evolution of the $R_\beta$-$R_S$ relationship (and the $t$-$(1-\eta)$ relationship), as well as the disparity $\Delta R$, as functions of photon energy $E_\gamma$ and agglomeration extent $\xi$.

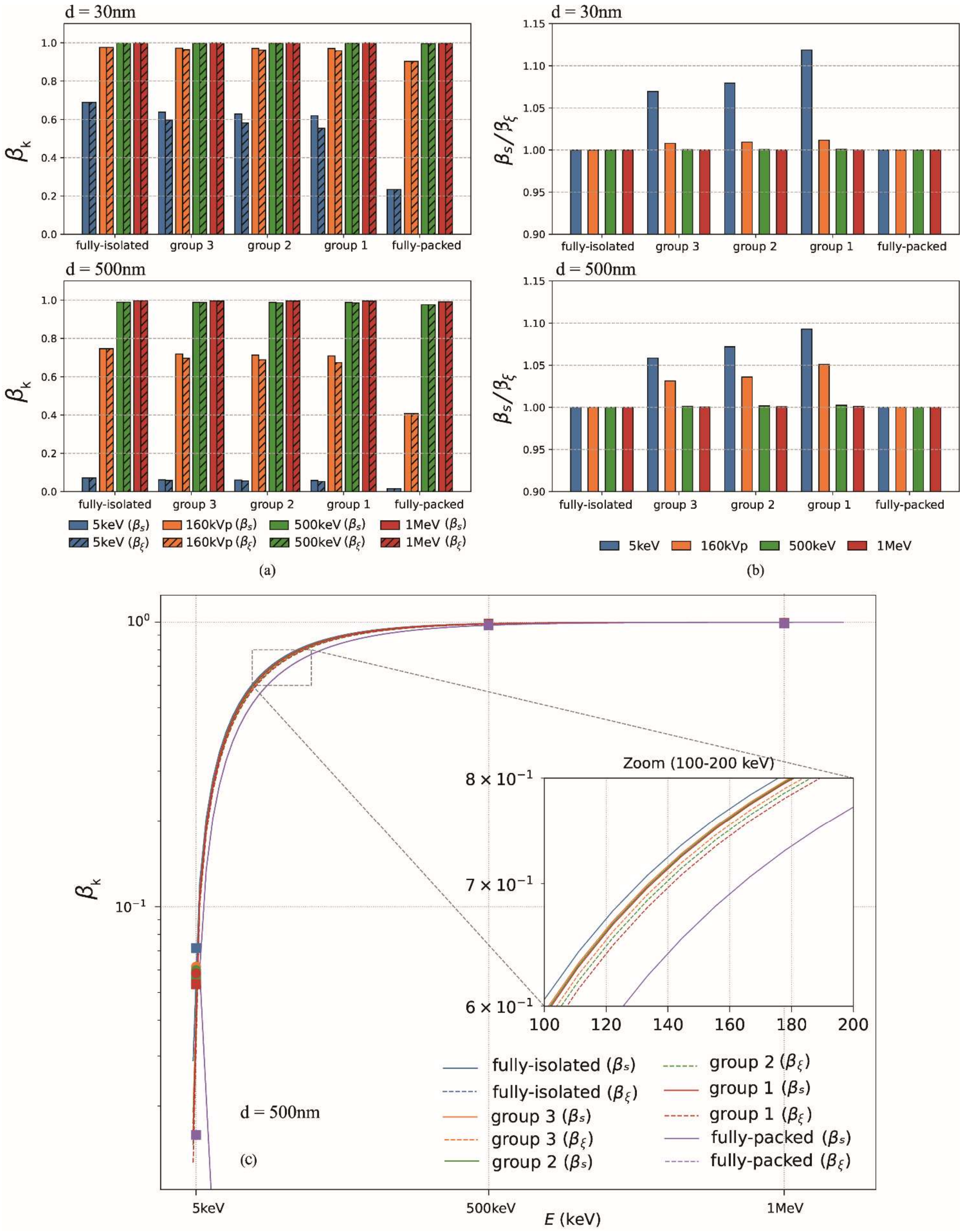

**Figure 4** Computational verification of the energy emission ratio disparity between nanoparticle systems and their fully isolated, equal-mass and equal-area single-sphere counterparts, together with the trends of this disparity across agglomeration extents ranging from fully isolated to fully packed states at photon energies of 5 keV, 160 kVp, 500 keV, and 1 MeV. (a) Comparison of $\beta_\xi$ and $\beta_S$ values for 30 nm and 50 nm nanoparticle systems; (b) Relative difference between $\beta_\xi$ and $\beta_S$ represented by the ratio $\beta_S/\beta_\xi$; (c) Comparison of the fitted $E_\gamma$-dependent $\beta_\xi$ and $\beta_S$ curves at different agglomeration extents for the 500 nm nanoparticle system.

### *2.5.6 Physical origin of the energy emission ratio disparity $\beta_k(\xi;E_\gamma)<\beta_k[R_S(\bar{S}(\xi));E_\gamma]$ — three potential mechanisms*

We now arrive at the central question underlying the above trends: what underlying factors make the energy emission ratio of nanoparticle systems consistently lower than that of fully isolated, equal-mass and equal-area single-sphere systems?

Fundamentally, the energy-emission ratio disparity can be attributed to three potential mechanisms in nanoparticle systems:

- Re-entry effect of emitted secondary electrons,
- Mass-deficiency and burial effects arising from inhomogeneous mass distribution within nanoparticle agglomerates,
- Path-tortuosity effect induced by interfacial scattering of secondary electrons.

Detailed analysis of these underlying physical mechanisms is deferred to the Discussion section.

### *2.5.7 Generalized expression, reduced forms, and practical prediction models*

For dual enhancement effects, the combined expression of the CPK-I and CPK-II models for LQ parameter modification is derived as equation (39), involving five parameters $K_s$, $K_\mathrm{m}$, $k$, $\eta$ and $t$. Based on the physical correlation analysis in the preceding sections, the general relationship between the two agglomeration parameters is fitted as $t=a(1-\eta)^b$, where $a$ and $b$ are photon-energy-dependent fitting parameters. Substituting $t=a(1-\eta)^b$ into equation (39) yields the expression with parameter degeneracy

$$\frac{\alpha''}{\alpha}=K_s s_0\bar{\lambda}^{\eta}+\bar{\lambda}K_\mathrm{m}m_0\beta_k\left(R_0\bar{\lambda}^{a(1-\eta)^b};\ E_\gamma\right),\quad \bar{\lambda}=A(1-e^{-kC}) \tag{49}$$

with four independent parameters $K_s$, $K_\mathrm{m}$, $k$ and $\eta$. However, this expression remains cumbersome for practical use.

From this general expression, reduced forms can be derived under particular conditions using different $t$–$(1-\eta)$ relations:

- Dual-effect agglomeration model: general relation $t=a(1-\eta)^b$

This formulation accounts for dual enhancement under agglomeration effects via the general $t$-$\eta$ relationship described by equation (49).

- Dual-effect agglomeration model: linearized relation $t=1-\eta$

As discussed in section 2.5.4, the inequality $\beta_\xi<\beta_S$ for the energy emission ratios of nanoparticle systems and their equal-$\bar{S}$ single-sphere counterpart directly yields a disparity between the equal-$\beta_k$ and equal-$\bar{S}$ single-sphere radii $R_\beta>R_\mathrm{S}$, and consequently $t>1-\eta$. When irradiation conditions drive the disparity $\Delta\beta_\mathrm{k}=\beta_S-\beta_\xi$ toward zero, it follows that $R_\beta\to R_\mathrm{S}$ and, accordingly, $t\to 1-\eta$.
This raises a question: what irradiation conditions may drive $\Delta\beta_\mathrm{k}$ to approach zero? From figure 3c, an intuitive deduction is that the photon energy, effective atomic number, and mass density of nanoparticles govern the applicability of $t=1-\eta$. Qualitatively, as photon energy increases, the energy emission ratios for nanoparticle systems and equal-$\bar{S}$ single-sphere systems both saturate towards unity, and $\Delta\beta_\mathrm{k}$ tends to zero. Higher photon energy drives $\Delta\beta_\mathrm{k}$ to approach zero, yielding $R_\beta\to R_\mathrm{S}$ and $t\to 1-\eta$. At fixed photon energy, the atomic number $Z_\mathrm{eff}$ and intrinsic density $\rho$ of nanoparticles govern the mean range and escape efficiency of secondary electrons, and thus the energy emission ratio $\beta_k(R;E_\gamma)$ from a single sphere of radius $R$. Nanoparticles with low $Z_\mathrm{eff}$ and low $\rho$ exhibit smaller $\Delta\beta_\mathrm{k}$, which also leads to $R_\beta\to R_\mathrm{S}$ and $t\to 1-\eta$.
The physical meaning of this approximation: when photon energy or material properties ($Z_\mathrm{eff}$, $\rho$) make the energy-emission ratios of nanoparticle assemblies and their equal-$\bar{S}$ single-sphere systems nearly identical, the equal-$\beta_k$ single-sphere radius $R_\beta$ declines and converges to the equal-$\bar{S}$ single-sphere radius $R_\mathrm{S}$, giving $R_\beta\to R_\mathrm{S}$ and accordlingly $t\to 1-\eta$. This linearized parameter relation $t=1-\eta$ reduces equation (49) to

$$\frac{\alpha''}{\alpha}=K_s s_0\bar{\lambda}^{\eta}+\bar{\lambda}K_\mathrm{m}m_0\beta_k\left(R_0\bar{\lambda}^{1-\eta};\ E_\gamma\right),\quad \bar{\lambda}=A(1-e^{-kC}) \tag{50}$$

with the four free parameters $K_s$, $K_\mathrm{m}$, $k$ and $\eta$ still retained.

- Dual-effect fully-isolated model: limiting behavior $t\to 0$, $(1-\eta)\to 0$

When agglomeration effects are negligible for both the surface-mediated biochemical enhancement and the local-dose and radiolysis enhancements mediated by mass-energy transfer of secondary electrons, the phenomenological agglomeration-characterizing parameters satisfy $t\to 0$ and $(1-\eta)\to 0$. Equation (49) reduces to dual-enhancement fully-isolated model, giving a two-parameter expression for $\alpha''/\alpha$ as a function of $C$, with the lumped constant $[K_s s_0+K_\mathrm{m}m_0\beta_k(R_0;E_\gamma)]$ and parameter $k$ (from equation (16))

$$\frac{\alpha''}{\alpha}=\left(K_s s_0+K_\mathrm{m}m_0\beta_k\left(R_0;\ E_\gamma\right)\right)\bar{\lambda}=\left(K_s s_0+K_\mathrm{m}m_0\beta_k\left(R_0;\ E_\gamma\right)\right)A(1-e^{-kC}) \tag{51}$$

- Dual-effect models: governing factors for relative strength

These dual-enhancement models above, with different $t$-$(1-\eta)$ relationships for varied agglomeration states, may exhibit marked differences in the relative strength between surface-mediated biochemical enhancement and mass-energy-transfer-mediated physicochemical enhancement. From $\bar{d}_1=\bar{d}_{1s}+\bar{d}_{1m}=K_s D s_0 \bar{\lambda}^{\eta-1}+K_m D m_0 \beta_k(R_0\bar{\lambda}^t)$ (equation (38)), the relative contributions of $\bar{d}_{1s}$ and $\bar{d}_{1m}$ to the equivalent dose $\bar{d}_1$ per nanoparticle are governed not only by agglomeration-dependent effective surafce area $s_1=s_0/\bar{\lambda}^{1-\eta}$ and mean energy emission $\bar{\varepsilon}_1=\psi D m_0 \beta_k(R_0\lambda^t)$ (equation (27-29)) per nanoparticle, but also by the material-dependent induction coefficients $K_s$ and $k_m$ (Table 1). For varying agglomeration extents, the coupled variation of $t$ and $1-\eta$ directly modulates the magnitudes of the effective surface area $s_1$ and mean energy emission $\bar{\varepsilon}_1$ per nanoparticle, but exert no influence on the two induction coefficients $K_s$ and $k_m$. Rather, to a large extent, the relative contributions of $\bar{d}_{1s}$ and $\bar{d}_{1m}$ to the equivalent dose $\bar{d}_1$, including cases of full dominance by a single effect, are governed by the relative magnitudes of the material-dependent induction coefficients.

- Reduction to single-effect models

When nanoparticle properties leasd to the dominance of surface-mediated biochemical enhancement, i.e., $K_s \gg k_m \to 0$, the general expression in equation (49) reduces to a single-effect agglomeration model for biochemical enhancement, yielding a three-parameter formulation of $\alpha''/\alpha$ as a function of $C$ (reproduced from equation (24))

$$\frac{\alpha''}{\alpha} = K_s s_0 \bar{\lambda}^{\eta}, \quad \bar{\lambda} = A(1-e^{-kC}) \tag{24}$$

with the lumped constant $A_s=K_s s_0 A^{\eta}$ and two independent constant parameters $k$ and $\eta$, both assumed invariant with $C$. When $1-\eta \to 0$, equation (24) further reduces to a single-effect fully-isolated model for biochemical enhancement, giving the two-parameter formulation for $\alpha''/\alpha$ versus $C$ below, with the reduced lumped constant $K_s s_0 A$ and parameter $k$.

$$\frac{\alpha''}{\alpha} = K_s s_0 \bar{\lambda} = K_s s_0 A(1-e^{-kC}) \tag{52}$$

Conversely, when radiosensitization is dominated by high-Z nanoparticle-mediated mass-energy transfer enhancement, i.e., $k_m \gg K_s \to 0$, the general expression of equation (49) reduces to single-effect agglomeration model for mass-energy transfer enhancement, yielding a formulation of $\alpha''/\alpha$ as a function of $C$ (reproduced from equation (37))

$$\frac{\alpha''}{\alpha} = K_m m_0 \bar{\lambda} \beta_k(R_0\bar{\lambda}^t; E_\gamma), \quad \bar{\lambda} = A(1-e^{-kC}) \tag{37}$$

with the lumped constant $K_m m_0$ and independent parameters $A$, $k$ and $t$. Both $k$ and $t$ are assumed invariant with $C$. It should be noted that the total degrees of freedom of equation (37) include the extrinsic structural parameters $K_m m_0, A, k, t,$ alongside the undetermined parameters embedded within $\beta_k(R; E_\gamma)$. A well defined set of independent parameters can only be obtained after an explicit analytical form of $\beta_k(R; E_\gamma)$ (e.g., equation (32)) is specified.

In the limit $t \to 0$, equation (37) further reduces to single-effect fully-isolated model for mass-energy-transfer enhancement, yieding a two-parameter formula for $\alpha''/\alpha$ as a function of $C$, with the lumped constant $K_m m_0 \beta_k(R_0; E_\gamma)A$ and parameter $k$

$$\frac{\alpha''}{\alpha} = K_m m_0 \beta_k(R_0; E_\gamma)\bar{\lambda} = K_m m_0 \beta_k(R_0; E_\gamma)A(1-e^{-kC}) \tag{53}$$

- Practical prediction models

The general model (equation (49)) using the nonlinear $t$-$\eta$ relation $t=a(1-\eta)^b$ and the approximate model (equation (50)) adopting the linearized $t=(1-\eta)$ still suffer from excessive undetermined parameters, limiting their practical applicability. For practical use, the above reduced equations (51), (24), (52), (37), (53) are grouped into three types of prediction models:

(a) CPK-iso — Unified fully-isolated model for dual- and single-effect enhancement

The dual-effect fully-isolated model described by equation (51), and the single-effect fully-isolated models for biochemical enhancement (equation (52)) and mass-energy transfer enhancement (equation (53)), adopts an identical linear formulation (differing only in coefficients) and can thus be uniformly expressed as

$$\alpha''/\alpha = A''(1-e^{-kC}) \tag{54}$$

where $A''$ and $k$ are the two free constant parameters. In practice, the parameters $(A'', k)$ in this formula can be determined from two sets of $(\alpha''/\alpha, C)$ data at two boundary concentrations $C_1$ and $C_2$, and then used for interpolation prediction (scheme described in Appendix E) of intermediate $\alpha''/\alpha$ values at concentrations satifying $C_1 < C < C_2$.

(b) CPK-I — Agglomeration model for surface-mediated biochemical enhancement

When equation (24) is used for the single biochemical enhancement effect, the lumped constant $A_s=K_s s_0 A^{\eta}$ and $k$ can be calculated from two sets of $(\alpha''/\alpha, C)$ data at two boundary concentrations $C_1$ and $C_2$, yet yield distinct values for different $\eta \in (2/3, 1)$. The optimal $\eta$ and corresponding $(A_s=K_s s_0 A^{\eta}, k)$ values should be determined via fitting to experimental data at intermediate concentrations, and then used to predict other intermediate $\alpha''/\alpha$ values. The interpolation-prediction method for the CPK-I model is described in Appendix E, as a refinement to the scheme reported in earlier study (Pan et al., 2022).

(c) CPK-II — Agglomeration model for mass-energy-transfer-mediated local-dose and radiolysis enhancement

When equation (37) is used for the single effect of mass-energy-transfer-mediated enhancement, substituting the calibrated single-sphere energy emission ratio $\beta_k(R; E_\gamma)=a(R+d)^b$ (equation (32)) into equation (37) yields

$$\frac{\alpha''}{\alpha} = K_\mathrm{m} m_0 \bar{\lambda} a\left(R_0 \bar{\lambda}^t + d\right)^b, \quad \bar{\lambda} = A(1-e^{-k\mathrm{C}}) \tag{55}$$

with the four independent parameters $(K_m m_0, A, k, t)$ unchanged. To perform interpolation prediction using equation (55), three parameters $(K_m m_0, A, k)$ remain undetermined after fixing $t$. Clearly, two sets of boundary (α″/α, C) data alone are insufficient to uniquely determine the full parameter set. Consequently, multiple experimental (α″/α, C) data sets spanning the valid concentration range are needed for constrained nonlinear least-squares fitting with physical bounds, to obtain the optimal set of $(K_m m_0, A, k)$ for each candidate $t$. After scanning all candidate values, the optimal $t$ value and the parameter set $(K_m m_0, A, k, t)$ can be identified by evaluating the prediction error. Such rigorous nonlinear fitting using equation (55), though theoretically feasible, remains challenging due to strong parameter correlation between $K_m m_0$ and $A$. For practical application, this model can be further simplified into a three-parameter form using a zero-order approximation of $\beta_k(R; E_\gamma)$. The derivation of this simplified formula and its range of applicability are discussed in Appendix C.

## 3. Experimental validation

### 3.1 Cell experiments

Human hepatocellular carcinoma (HepG2) cells were used for in vitro model validation of the radiosensitization mediated by polyethylene glycol-coated gold nanoparticles (PEG-GNPs). Detailed procedures for cell culture, cytotoxicity assay, cell irradiation and clonogenic survival assay are provided in Appendix D. In vitro validations of the practical models for interpolation prediction of the LQ parameter correction α”/α were performed using the clonogenic dose-survival data from these PEG-GNP radiosensitization experiments with HepG2 cells.

### 3.2 Model validation

*3.2.1. Fully-isolated model: CPK-iso validation*

Cytotoxicity assays identified a safe PEG-GNP concentration range with an upper limit of 2 μg/ml for validating the unified fully-isolated model, CPK-iso. The interpolation-based prediction was performed using the boundary (α”/α, C) data at two concentrations ($C_1$=0.04 μg/ml, $C_2$=2 μg/ml). Intermediate α”/α values at two concentrations (0.2 μg/ml, 0.4 μg/ml) were predicted via the unified formula given by equation (54), and the predicted LQ survival curves were compared against the clonogenic dose-survival data of the sensitized cells at the same concentrations. The prediction error (relative deviation) between the predicted and experimental survival fractions at each concentration was calculated as

$$\text{deviation} = \frac{\left(SF_\text{pred} - SF_\text{exp}\right)}{SF_\text{exp}} \tag{56}$$

The experimental dose-survival data are expressed as mean ± SD; therefore, error propagation is included in the calculation of the prediction error. For equation (56), the uncertainty (standard deviation) of the relative deviation is calculated as

$$\sigma(\text{deviation}) = \frac{SF_\text{pred}}{SF_\text{exp}^2}\sigma\left(SF_\text{exp}\right) \tag{57}$$

Note that equation (57) only accounts for the uncertainty in experimental $SF_\text{exp}$, while the uncertainty of $SF_\text{pred}$ originating from fitted model parameters is neglected as a simplifying assumption. The calculated $\sigma$ (deviation) should be regarded as a lower estimate of the total uncertainty.

*3.2.2 Agglomeration models: CPK-I and CPK-II validation*

The CPK-I model for isolated surface-mediated biochemical enhancement (equation (24)) was tested to predict α”/α over the PEG-GNP concentration range of 0.04-2 μg/ml. Boundary (α″/α, C) data were used to compute α”/α at two intermediate concentrations (0.2 μg/ml, 0.4 μg/ml) via equation (24) for $\eta \in (2/3, 1)$, to determine the optimal $\eta$ and the parameter set $(A_\mathrm{s}=K_s s_0 A^\eta, k, \eta)$. The fitting procedure for optimal $\eta$ by goodness-of-fit evaluation is detailed in Appendix E. Using the identified parameters $(A_\mathrm{s}=K_s s_0 A^\eta, k, \eta)$, cell survival at the intermediate concentrations was re-predicted for a preliminary validation (further validations using independent concentrations are still needed). Prediction errors were compared against those from the CPK-iso model ($\eta$=1).

The CPK-II model for the isolated mass-energy-transfer-mediated enhancement was validated over the same concentration range, using equation (C2) in Appendix C — the simplified form of equation (55)). This formulation originates from the power-function approximation of the single-sphere energy-emission ratio $\beta_k(R; E_\gamma)$ (equation (C1)) and reduces CPK-II to a three-parameter formula characterised by the parameter set $(A_\mathrm{m}, k, t)$. Boundary (α″/α, C) data and a grid search over $t$ were used to predict α”/α at the two intermediate concentrations (0.2μg/ml, 0.4 μg/ml). The goodness-of-fit analysis for the optimal $t$ follows the same fitting scheme as for the optimal $\eta$ in CPK-I (Appendix E). With the determined parameter set $(A_\mathrm{m}, k, t)$, cell survival at the two concentrations was re-predicted for preliminary validation (further validations using independent concentrations are still needed). Prediction errors were compared with those from the CPK-iso model ($\eta$=1).

### 3.3 Results

#### *3.3.1. Radiosensitization of HepG2 cells*

The cytotoxicity assay showed that the relative viability of HepG2 cells remained above 0.9 (
Figurea) for the PEG-GNP concentrations below 2 μg/mL. Therefore, 2 μg/mL was selected as the safe upper-limit PEG-GNP concentration for the *in vitro* radiosensitization experiments with HepG2 cells. Survival fractions of irradiated HepG2 cells pretreated with different PEG-GNP concentrations (0, 0.04, 0.2, 0.4, 2 μg/mL) were obtained from the clonogenic assay, and the corresponding dose survival curves are presented in
Figureb. The PEG-GNPs mediated evident radiosensitization in HepG2 cells with a clear concentration dependence. At each concentration, the dose survival data were fitted to the LQ model, and the resulting LQ parameter (α,β) are summarized in Table . The relative correction α″/α at the boundary concentrations (0.04μg/ml, 2 μg/ml) was also calculated for interpolation prediction.

#### *3.3.2. Validation of the proposed prediction models*

For the unified fully-isolated model CPK-iso, the experimental α″/α data at the boundary concentrations (0.04, 2 μg/mL) (Table 4) were used to predict the α″/α values at the two intermediate concentrations (0.2,0.4 μg/mL) via equation (54). The fitted parameters ($A$”, $k$) and the predicted LQ parameters (α, β) are given in Table 5. The predicted LQ survival curves are plotted in figures 6a and 6b, together with the corresponding experimental data at the two intermediate concentrations for comparison. For each concentration, the prediction error was computed using equations (56) and (57), and the results are presented in figures 6c and 6d.

For the single-effect agglomeration models, variations in $\eta$ (CPK-I) and $t$ (CPK-II) lead to different α″/α predictions from each model, resulting in distinct LQ survival curves at the same intermediate concentrations. The goodness-of-fit between the predicted and experimental LQ survival data was evaluated using the coefficient of determination ($R^2$) by equation (E1) (detailed in Appendix E). For the CPK-I model, the $\eta$-dependent variation of the goodness-of-fit value $R^2$ is presented in
Figure **E1**, from which the best $\eta$ value was identified at $\eta$≈0.925115. The corrersponding fitting parameters ($A_s$, $k$) of the α″/α expression (equation (24)) and the predicted (α, β) values are given in Table 5. The re-predicted LQ survival curves and the prediction errors at the two intermediate concentrations are presented in figure 7 alongside those obtained from the fully-isolated model (CPK-iso) for comparison. For the CPK-II model (using its simplified three-partameter form), the $t$-dependent variation of the $R^2$ value is given in
Figure **E1**, from which the best $t$ was identified at $t \approx 0.061980$. The fitting parameters ($A_m$, $k$) for the α″/α expression (equation (C2)) and the predicted (α, β) values are shown in Table 5. The re-predicted survival curves and prediction errors at the same intermediate concentrations are given in figure 8, together with those from the CPK-iso and CPK-I models for comparison. An overall comparison of the prediction performance for the three models is provided in section 3.3.3.

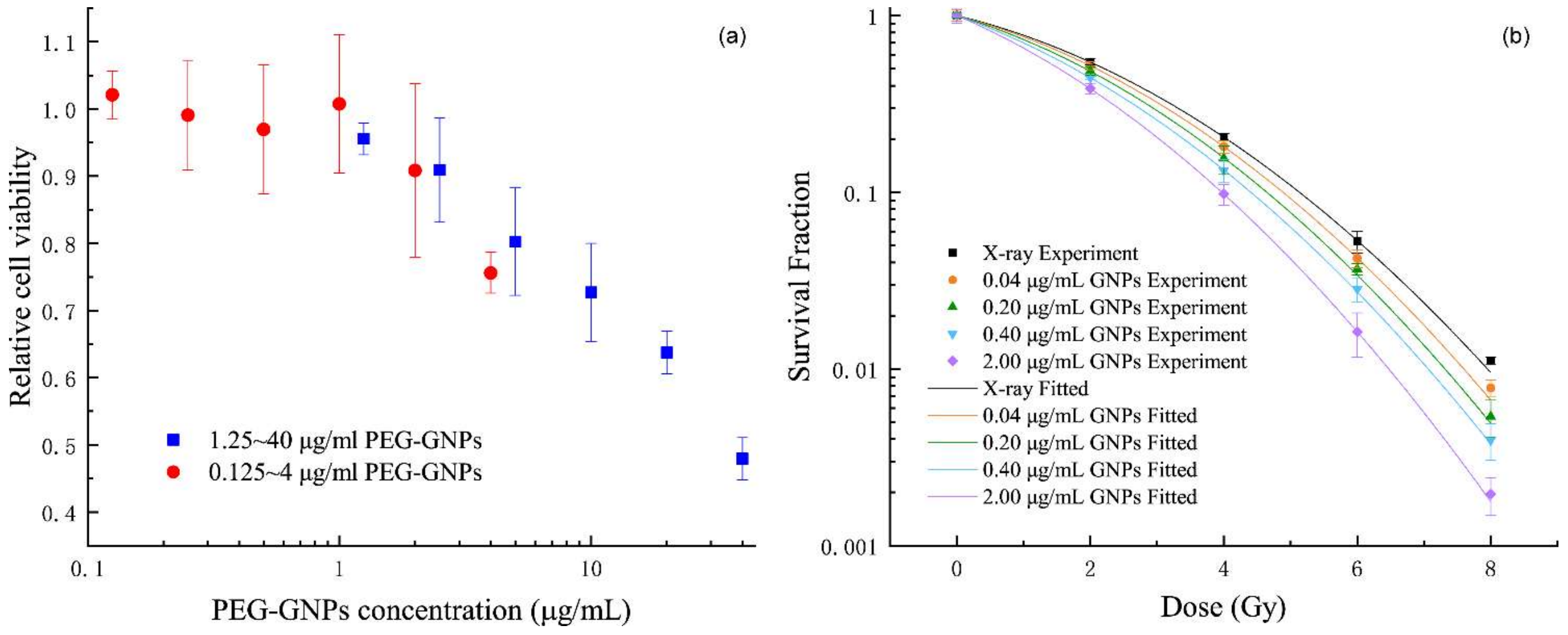

**Figure 5.** (a) Cytotoxicity assay of the PEG-GNPs used for HepG2-cell radiosensitization experiments, showing relative cell viability to determine safe working concentration range; (b) Clonogenic dose-survival curves of the PEG-GNP-sensitized HepG2 cells at different concentrations under 160 kVp X-ray irradiation.

**Table 4.** LQ parameters modification: measured for the clonogenic dose survival of HepG2 cells at various concentrations of PEG-GNPs.

| Concentration (μg /mL) | α Experimental ($Gy^{-1}$) | β Experimental ($Gy^{-2}$) | α″ (mean) ($Gy^{-1}$) | α″/α (mean) |
|---|---|---|---|---|
| 0 (Control) | 0.21057 ± 0.00177 | 0.04633 ± 5.61213E-4 | 0 | 0 |
| 0.04 | 0.22777 ± 0.00139 | 0.0498 ± 4.59094E-4 | 0.0172 | 0.08168 |
| 0.20 | 0.26252 ± 0.00433 | 0.05017 ± 0.00149 | | |
| 0.40 | 0.30786 ± 0.00198 | 0.04871 ± 7.02513E-4 | | |
| 2.00 | 0.37140 ± 0.00035 | 0.05253 ± 1.36843E-4 | 0.1608 | 0.76378 |

**Table 5.** LQ parameter modification: predicted using the CPK-iso, CPK-I and CPK-II models at the two intermediate concentrations.

| Model | Fitting Parameters | Concentration (μg/mL) | α Predicted ($Gy^{-1}$) | β (Control) ($Gy^{-2}$) |
|---|---|---|---|---|
| CPK-iso | $A''$= 0.76652 | 0.20 | 0.28009 | 0.04633 |
| | $k$ = 2.81699 | 0.40 | 0.31967 | 0.04633 |
| CPK-I | $A_s$ = 0.77077 | 0.20 | 0.27532 | 0.04633 |
| $\eta$ = 0.925115 | $k$ = 2.31300 | 0.40 | 0.31230 | 0.04633 |
| CPK-II | $A_m$ = 0.76817 | 0.20 | 0.27770 | 0.04633 |
| $t$ = 0.061980 | $k$ = 2.56333 | 0.40 | 0.31604 | 0.04633 |

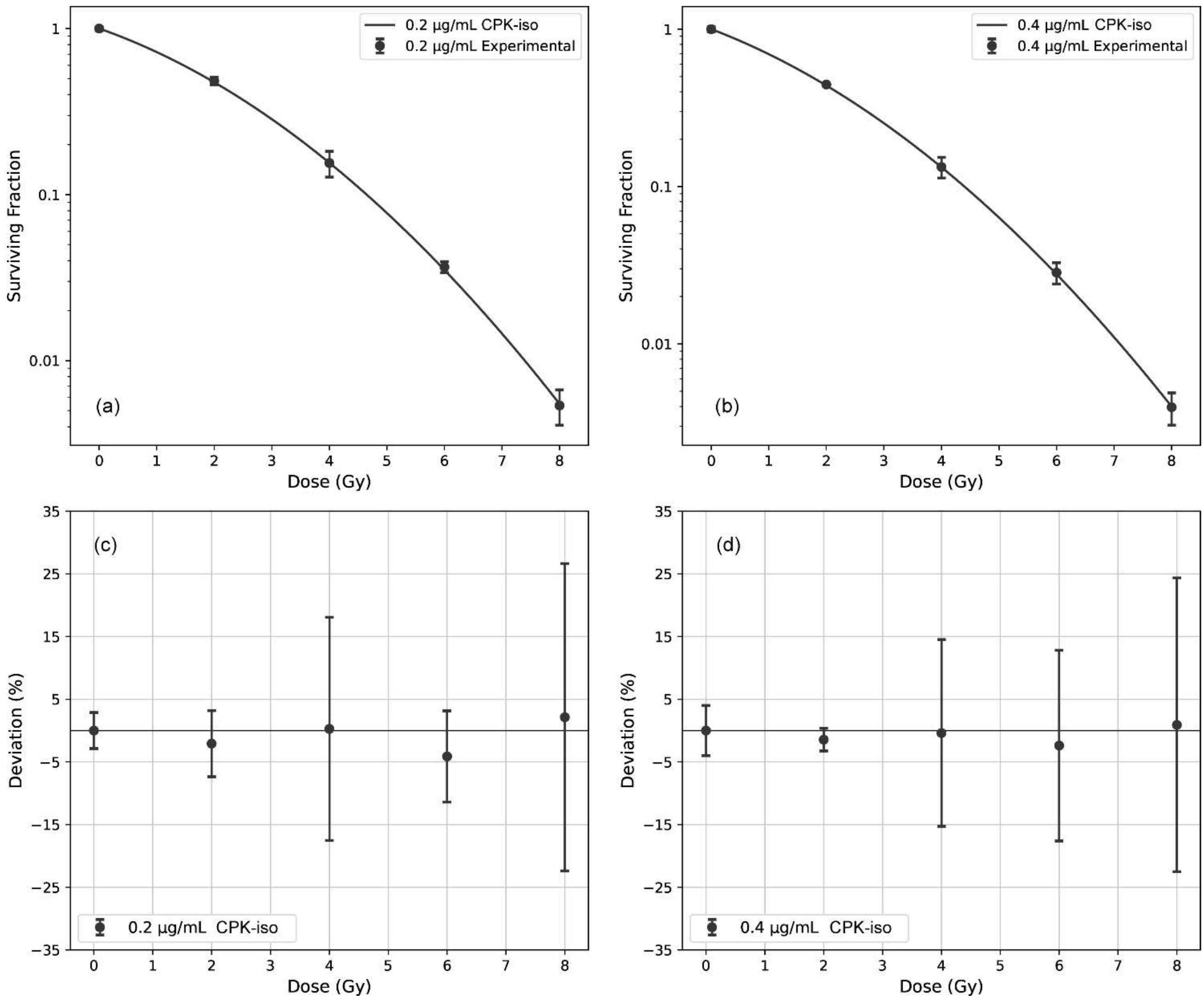

**Figure 6.** In vitro validation of the CPK-iso model using PEG-GNP-mediated radiosensitization of HepG2 cells. (a,b) Comparison of LQ survival curves (solid lines) predicted by the CPK-iso model with clonogenic survival data (solid squares) measured at intermediate PEG-GNP concentrations (0.2, 0.4 µg/ml) within the 0.04–2 µg/ml range, under 160 kVp X-ray irradiation at a dose rate of 0.75 Gy/min. (c,d) Prediction errors (relative deviation) of the CPK-iso model relative to clonogenic survival data, demonstrating that the relative deviation remains below ±5%.

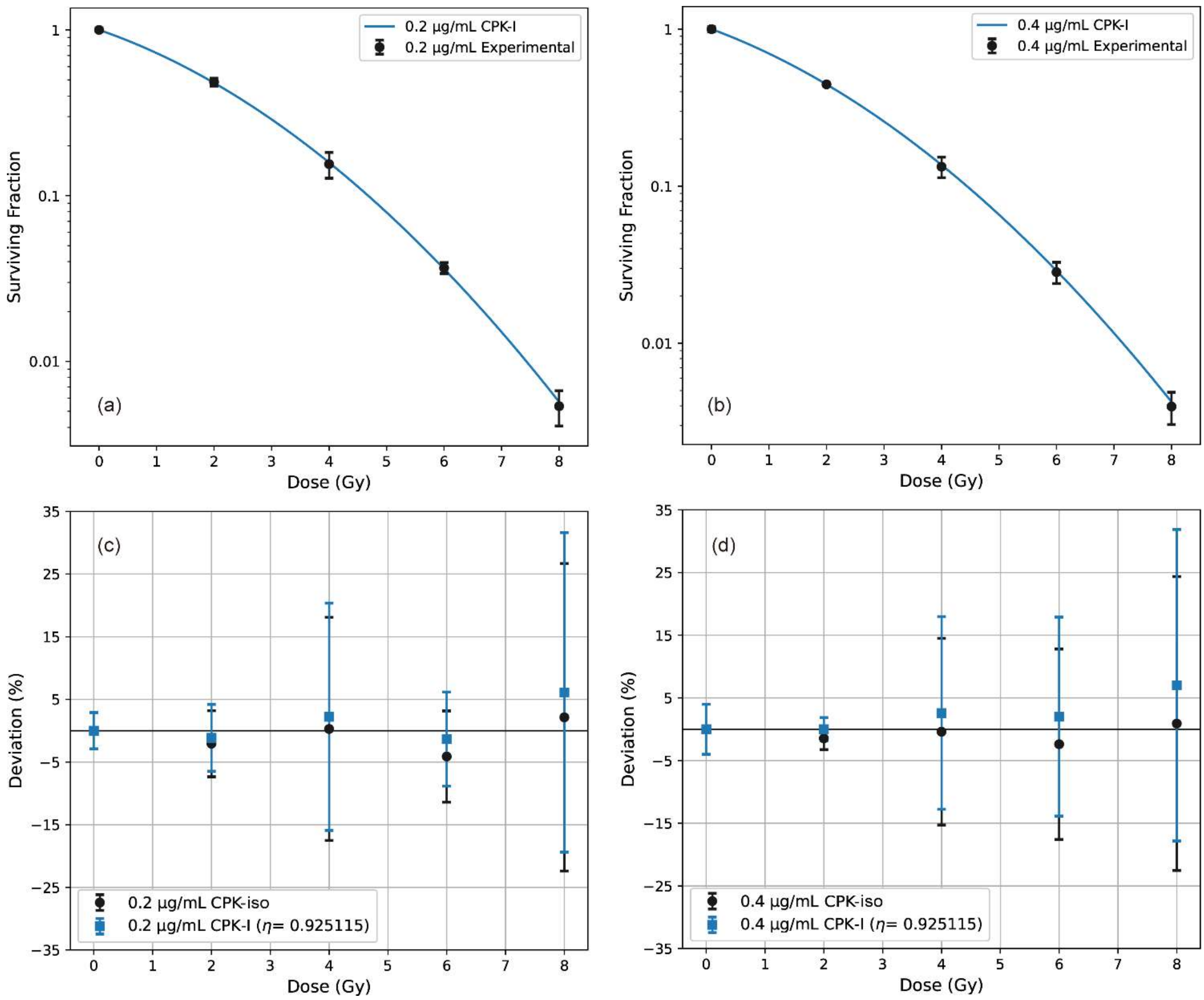

**Figure 7**. In vitro validation of the CPK-I model using PEG-GNP-mediated radiosensitization in HepG2 cells. (a-b) Comparison of LQ survival curves (blue solid lines) predicted by the CPK-I model against clonogenic survival data (black solid squares) measured at two intermediate PEG-GNP concentrations (0.2, 0.4 μg/ml) within the 0.04–2 μg/ml range, under 160 kVp X-ray irradiation at a dose rate of 0.75 Gy/min. (c-d) Prediction errors (relative deviation) of the CPK-I model (blue solid squares, $\eta$=0.92515) relative to the clonogenic survival data, demonstrating that its relative deviation falls within ±7%. Shown for comparison are the prediction errors of the CPK-iso model (black solid circles), with a relative deviation below ±5%.

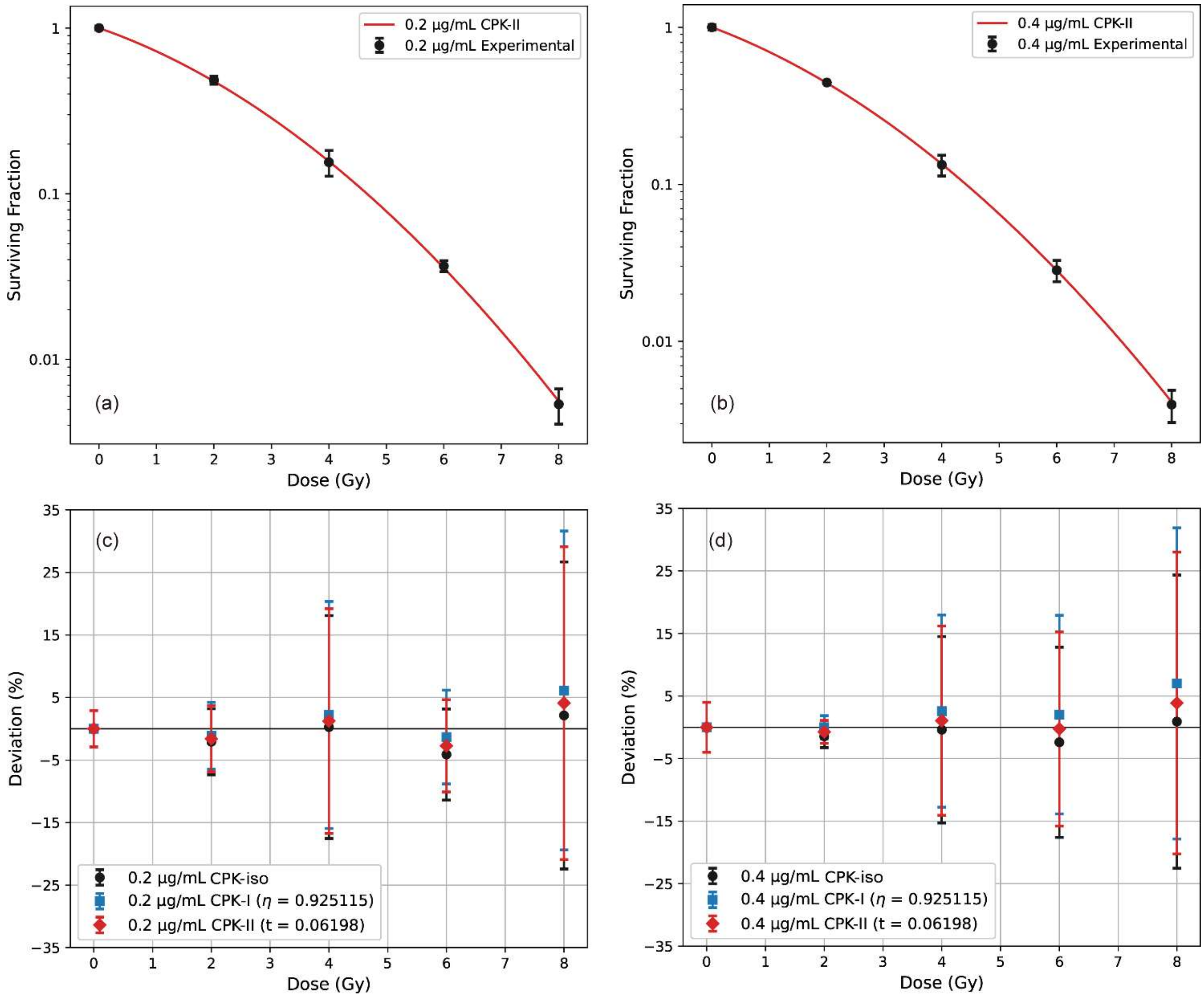

**Figure 8**. In vitro validation of the CPK-II model using PEG-GNP-mediated radiosensitization in HepG2 cells. (a-b) Comparison of LQ survival curves (red solid lines) predicted by the CPK-II model against clonogenic survival data (black solid squares) measured at two intermediate PEG-GNP concentrations (0.2, 0.4 μg/ml) within the 0.04–2 μg/ml range, under 160 kVp X-ray irradiation at 0.75 Gy/min. (c-d) Prediction errors of the CPK-II model (red solid squares, $t$=0.06198) relative to clonogenic survival, indicating a relative deviation within ±5%. The prediction errors of the CPK-iso model (black solid circles, within ±5%) and those of the CPK-I model (black solid circles, within ±7%) are also presented for comparison.

*3.3.3 Overall comparison of the prediction performance*

For the predicted LQ survival and corresponding errors obtained by the three models at the two intermediate concentrations (figures 6-8), four statistical metrics, namely root-mean-square deviation (RMSD), mean-absolute deviation (MAD), mean deviation (MeanDev) and standard deviation of prediction error (SDerr), were adopted to quantitatively compare the overall prediction performance of the CPK-iso, CPK-I and CPK-II models across the full dose range of 0–8 Gy. The RMSD was employed to assess the global prediction accuracy. The MAD was used to quantify the average error magnitude. The SDerr reflected the fluctuation and stability of residuals, while the MeanDev indicated whether systematic overestimation or underestimation exists in the predictions, serving as an indicator of systematic bias for each model. Computational formulas for these statistical metrics are provided in Appendix F.

**Table 6**. Statistical metrics (RMSD, MAD, SDerr, MeanDev) for evaluating the prediction performance of the CPK-iso, CPK-I, and CPK-II models at PEG-GNP concentrations of 0.2 μg/mL and 0.4 μg/mL over the dose range of 0–8 Gy.

| C = 0.2 μg/mL | CPK-iso | CPK-I | CPK-II |
|---|---|---|---|
| RMSD (%) | 2.28 | 3.01 | 2.38 |
| MAD (%) | 1.72 | 2.16 | 1.94 |
| SDerr (%) | 2.40 | 3.10 | 2.66 |
| Mean Dev (%) | -0.76 | 1.17 | 0.19 |
| C = 0.4 μg/mL | CPK-iso | CPK-I | CPK-II |
| RMSD (%) | 1.32 | 3.47 | 1.83 |
| MAD (%) | 1.03 | 2.33 | 1.18 |
| SDerr (%) | 1.28 | 2.88 | 1.84 |
| Mean Dev (%) | -0.66 | 2.33 | 0.79 |

Table 6 summarizes the four prediction-error metrics (RMSD, MAD, SDerr, and MeanDev) for the three models at the two intermediate validation concentrations (0.2 and 0.4 μg/mL). CPK-iso achieves the lowest values of RMSD (2.28 % and 1.32 %) and MAD (1.72 % and 1.03%) at both concentrations, demonstrating the highest overall prediction accuracy. Its SDerr (2.40% and 1.28 %) is also the lowest among the models, indicating minimal random fluctuation in prediction results. CPK-II yields slightly higher RMSD (2.38 % and 1.83 %), MAD (1.94 % and 1.18 %) and SDerr (2.66 % and 1.84 %) values than CPK-iso, yet substantially lower values than CPK-I, indicating acceptable precision and error stability. In contrast, CPK-I yields markedly elevated RMSD (3.01% and 3.47%), MAD (2.16 % and 2.33 %), and SDerr (3.10 % and 2.88 %) values at both the validation concentrations, corresponding to poorer accuracy and larger scattering of residuals. For all three models, the SDerr value at 0.4 μg/mL is lower than that at 0.2 μg/mL, indicating a more stable performance at higher validation concentration.

Regarding the systematic bias quantified by MeanDev: CPK- iso yields mild negative deviations (−0.76 % and −0.66 %), indicating a slight tendency to underpredict survival fractions. CPK-II exhibits very small positive mean deviations (0.19 % and 0.79 %) corresponding to nearly unbiased predictions. CPK-I shows consistently positive mean deviations (1.17 % and 2.33 %), with the bias increasing with concentration, pointing to a concentration-dependent overprediction trend.

Taken together, the unified fully-isolated model, CPK- iso, delivers the most accurate and stable interpolation predictions with low random error fluctuation; the single-effect agglomeration model for mass-energy-transfer mediated enhancement, CPK-II, yields satisfactory prediction performance with negligible bias, implying a more physically reasonable formulation; the single-effect agglomeration model for surface-mediated biochemical enhancement, CPK-I, exhibits reduced accuracy and systematic overestimation bias under the experimental conditions. Overall, for the cell-irradiation experiment, across the 0–8 Gy dose range, all three models offer sufficiently high accuracy. Nevertheless, the rigorous quantitative comparison based on RMSD, MAD and SDerr still reveals a performance hierarchy of CPK-iso>CPK-II > CPK-I, while the comparison based on mean deviation ranks the systematic-bias performance as CPK-II > CPK-iso > CPK-I.

## 4. Discussion

In constructing the general model for dual enhancement, analysis of the physical correlation between the two agglomeration parameters $\eta$ and $t$, separately introduced in the CPK-I model for surface-biochemical enhancement and the CPK-II model for mass-energy-transfer mediated physicochemical enhancement, raises a question of physical significance: why does a randomly agglomerated nanoparticle system consistently exhibit a smaller secondary-electron energy-emission ratio than its equal-mass, equal-area counterpart consisting of fully-isolated monodisperse single spheres, for all photon energies and agglomeration extents ?

Here we propose three underlying mechanisms intrinsic to agglomerated nanoparticle systems that account for this disparity:

(1) Mass inward-shift effect for secondary-electron emission arising from nanoparticle agglomeration

Let $\Omega_S$ be the reference system containing $N$ fully isolated monodisperse spheres, and $\Omega_A$ the corresponding agglomerated nanoparticle assembly, subjected to the dual constraints of equal total mass (equal volume $V$ for uniform density) and equal exposed surface area $S$. For the nanoparticle assembly, the inter-particle contact surfaces are sealed internally and no longer serve as electron-escape boundaries. To satisfy the equal-area constraint, concave gullies and indentations are inevitably formed within the agglomerate topology.

To characterize the spatial positional relationship between internal mass and external electron-escape boundaries, a physical field function named shortest secondary-electron escape distance can be defined as

$$d(\boldsymbol{r}) = \min_{s \in \partial\Omega_{\text{out}}} \parallel \boldsymbol{r} - \boldsymbol{s} \parallel \tag{58}$$

where $\boldsymbol{r}$ is the position vector of an arbitrary infinitesimal mass element in the nanoparticle domain, $\boldsymbol{s}$ is the position vector of any point on the electron-escape surface $\partial\Omega_{\text{out}}$, and $\parallel \boldsymbol{r} - \boldsymbol{s} \parallel$ is the straight-line distance between the mass element and the escape point. The value of $d(\boldsymbol{r})$ physically indicates the minimum transport distance that secondary electrons generated at position $\boldsymbol{r}$ need to travel to escape from the nanoparticle system into the external medium.

To statistically quantify the overall spatial distribution of the system mass, the mass-weighted mean escape distance $\langle d \rangle$ is introduced, with the expression

$$\langle d \rangle = \frac{1}{V} \iiint_{\Omega} d(\boldsymbol{r}) \mathrm{d}V \tag{59}$$

where $V$ is the total volume of the nanoparticle system. Given the uniform density of the nanoparticles, the total volume $V$ is proportional to the total mass, guaranteeing the validity of $\langle d \rangle$ for mass-weighted statistics. A larger $\langle d \rangle$ indicates that the system mass is statistically distributed farther from the external surface, i.e., a mean inward shift of the mass distribution.

In the reference system $\Omega_S$ of fully isolated spheres, each sphere has a standard convex geometry. The shortest escape path of all internal mass elements is a radial straight line pointing to its own spherical surface, without any topological obstacles or extended transport paths. For a single sphere of radius $r_i$, the mean escape distance is $\langle d \rangle_i = r_i/4$. The overall mass-weighted mean escape distance for the mutually isolated monodisperse sphere system is derived as $\langle d \rangle_S = \frac{1}{V}\sum_{i=1}^{N} V_i \frac{r_i}{4} = \frac{r_i}{4}$, which provides a rigorous and stable baseline for mass distribution statistics.

In contrast, the agglomerated nanoparticle system $\Omega_A$ exhibits distinct topological anomalies (concave gully structures and protruding thin-wall structures) under the equal-mass and equal-area constraints. The concave gully structures significantly extend the shortest escape distance of deep internal mass elements, because the originally adjacent electron-escape surfaces of primary nanoparticles are sealed by adhesion, forcing secondary electrons originating from deep mass regions to travel through long and tortuous gully channels to reach the valid outward-facing escape surfaces. This leads to a sharp increase in $d(\boldsymbol{r})$ for buried mass regions, namely $d_{\text{A}}(\boldsymbol{r}_{\text{deep}}) > d_{\text{S}}(\boldsymbol{r}_{\text{deep}})$. Although the local protruding thin-wall structures endow a small fraction of shallow mass with tiny electron escape distances, their volume fraction and mass weight are too small to offset the statistical rise in the mean escape distance arising from deeply buried mass. Integration over the total mass yields the core inequality for mass-weighted mean escape distances $\langle d \rangle_A > \langle d \rangle_S$, indicating that the inhomogeneous agglomeration of nanoparticles induces an obvious mean inward shift of the mass distribution relative to the monodisperse single-sphere system.

For secondary electrons with a maximum escape range $L_{\max}$ inside the solid, only the mass elements satisfying $d(\boldsymbol{r}) \le L_{\max}$ can generate escapable secondary electrons. The fraction of effective escape mass is expressed as $F_{\text{esc}} = \frac{1}{V}\iiint_{\Omega, d(\boldsymbol{r}) \le L_{\max}} \mathrm{d}V$. The inward shift of the mass distribution increases the overall escape distance of agglomerated nanoparticle system, reduces the proportion of shallow mass within the secondary-electron effective escape range, and ultimately leads to $F_{\text{esc},A} < F_{\text{esc},S}$, which fundamentally explains the reduced secondary-electron energy emission ratio of nanoparticle agglomerates.

(2) Path-tortuosity effect on secondary-electron emission arising from solid-solid interfacial scattering
In addition to the mass inward-shift effect intrinsic to randomly packed nanoparticles, a path-tortuosity effect arising from solid-solid interface scattering inside agglomerates further impairs secondary-electron emission. For isolated monodisperse spheres, secondary-electron transport from the generation site to the escape surface occurs entirely within a homogeneous single solid medium and the travel path coincides with the straight-line distance. By contrast, for electrons originating from deep-seated mass within nanoparticle agglomerates, reaching an unobscured escape surface is subject not only to the mass inward-shift effect that enlarges geometric escape distances but also frequently requires traversing numerous inter-particle solid-solid interfaces. Even if the straight-line distance is short, scattering takes place each time secondary electrons pass across nanoparticle contact zones or inter-particle gaps. Each crossing of a particle boundary raises the tortuosity of electron trajectories. Consequently, the effective transport distance (actual path length) for electrons produced in buried regions of agglomerates exceeds the geometric straight-line distance, which further exacerbates secondary-electron attenuation.

(3) Re-entry effect of escaping electrons due to solid-angle confinement on the escape surface of agglomerates
For fully-isolated single spheres, nearly all secondary electrons emitted from unobscured surfaces can escape and contribute fully to the total energy-emission ratio. Each surface element of the isolated convex sphere provides a full solid-angle of $2\pi$ pointing toward the surrounding medium.
By contrast, in nanoparticle agglomerates, many geometrically unobscured surface facets lie within inter-particle gaps and concave cavities, with surface normals pointing toward internal voids or neighboring nanoparticles. Although the effective exposed area of nanoparticle agglomerates is defined as the line-of-sight surface visible to infinity and matches that of the reference fully-isolated sphere system, the topological features greatly reduce the quality of these electron-escaping outlets, because secondary-electron emission is not strictly normal to the surface but exhibits energy-dependent angular anisotropy. Electrons emitted at grazing or lateral angles, even after passing through the effective surface, may travel across nanoscale inter-particle gaps and be recaptured by adjacent particles.
Consequently, an equal effective area does not guarantee equivalent electron escape efficiency. Nanoparticle agglomerates impose a limited solid-angle and lower free-path probability for electron escape, forming an extra internal energy-loss channel absent in fully-isolated single-sphere systems.

As for the trends of the energy-emission ratio disparity between nanoparticle system and its fully isolated equal-area single-sphere counterpart, the three mechanisms also provide a physical interpretation of how $\Delta\beta_k$ depends on photon energy and agglomeration state. Figure 3c shows that, at a given agglomeration degree of a nanoparticle system, as photon energy rises, $\Delta\beta_k$ first increases, reaches maximum at a specific energy, and then decreases and approaches zero. At fixed photon energy, as the agglomeration degree increases from fully isolated state toward a fully packed single sphere, $\Delta\beta_k$ rises from zero to a maximum value and subsequently decays back toward zero. The evolution trends of $\Delta\beta_k$ are physically governed by the variations of the three effects: for nanoparticle systems at a fixed agglomeration degree, as photon energy increases, the mass inward-shift, path-tortuosity and re-entry effects emerge, grow with photon energy and the associated electron range, peak, and are progressively suppressed toward zero as photon energy and secondary-electron range approach infinity. At a given photon energy and fixed secondary-electron range, as nanoparticle agglomeration increases from fully-isolated state to fully-packed single sphere, the magnitude of the three effects rises from zero for fully-isolated nanoparticles, reaches a maximum at intermediate agglomeration, and falls back to zero for the fully-packed single-sphere. In short, the evolution of the three effects directly accounts for the trends of the energy-emission ratio disparity $\Delta\beta_k$ between nanoparticle systems and their equal-mass and equal-area counterparts of fully-isolated monodisperse single spheres.

## 5. Conclusion

In this study, the compound Poisson killing model for nanoparticle radiosensitization was extended from the surface-effect model to a framework incorporating both surface-mediated biochemical enhancement and mass-energy-transfer-mediated local-dose and radiolysis enhancement. The equivalent dose per nanoparticle for the two enhancement channels is governed by the agglomeration extent of nanoparticle systems via their effective area and energy emission ratio, and is equivalently formulated using the radius dependences of the effective area and energy emission ratio of two monodisperse, equal-mass, fully isolated single-sphere systems, matched respectively in total effective area and energy emission ratio. In constructing these radius-dependent functions, two phenomenological factors, $t$ and $\eta$, were introduced to characterize the agglomeration extent. Based on the expressions for the equivalent dose per nanoparticle, the relative corrections $\alpha''/\alpha$ to the LQ parameters were derived for each enhancement channel under the assumption of additive compound Poisson killing. Their combination produced a generalized dual-enhancement expression. To reduce the number of free parameters, the correlation between $t$ and $\eta$ was established through physical interpretation of the energy-emission-ratio disparity between nanoparticle systems and their equal-area single-sphere counterparts. The parameter-reduced expression yielded three prediction models for $\alpha''/\alpha$: CPK-iso (fully-isolated model for dual- and single-effect enhancement), CPK-I (agglomeration model for surface-mediated biochemical enhancement), and CPK-II (agglomeration model for mass-energy-transfer-mediated local-dose and radiolysis enhancement).

Validation against the clonogenic dose-survival data of PEG-GNP-exposed HepG2 cells (0.04-2 μg/mL) showed that CPK-iso achieved the highest prediction accuracy and stability, with errors within ±5%, and that CPK-II performed satisfactorily with negligible systematic bias, while CPK-I exhibited larger deviations and concentration-dependent overestimation, as anticipated because this model, constructed for surface-mediated biochemical enhancement, cannot adequately capture the secondary-electron mass-energy transfer that governs high-Z GNP radiosensitization.

By physically decoupling the two enhancement mechanisms, this extended model provides a mechanism-based quantitative framework for predicting clonogenic survival in high-Z nanoparticle radiosensitization.

## Acknowledgment

The work was supported by National Natural Science Foundation of China under Grant number 12275056 and 12075063.

## Appendix A. Radius-dependence calibration of single-sphere energy-emission ratio

The simulated radius-dependence of the gold-sphere energy-emission ratio $\beta_k$ was well-fitted by equation (32). The fitting parameters are given in Table A1. The simulated data and corresponding curves are given in Figure A1.

**Table A1**. Fitting parameters for equation (32) describing the radius-dependence of gold-sphere energy-emission ratio $\beta_k$

| Energy | $a$ | $b$ | $d$ | $R^2$ | Adjusted_$R^2$ | RMSE |
|---|---|---|---|---|---|---|
| 5 keV | 133.5503 | -1.36024 | 33.68395 | 0.997929 | 0.997239 | 0.015172 |
| 160 kVp | 37.15235 | -0.60443 | 397.1444 | 0.999702 | 0.999603 | 0.002281 |
| 500 keV | 1.028903 | -0.00686 | 63.26196 | 0.999366 | 0.999154 | 0.000127 |
| 1 MeV | 1.01018 | -0.00248 | 61.04885 | 0.996865 | 0.99582 | 0.000104 |

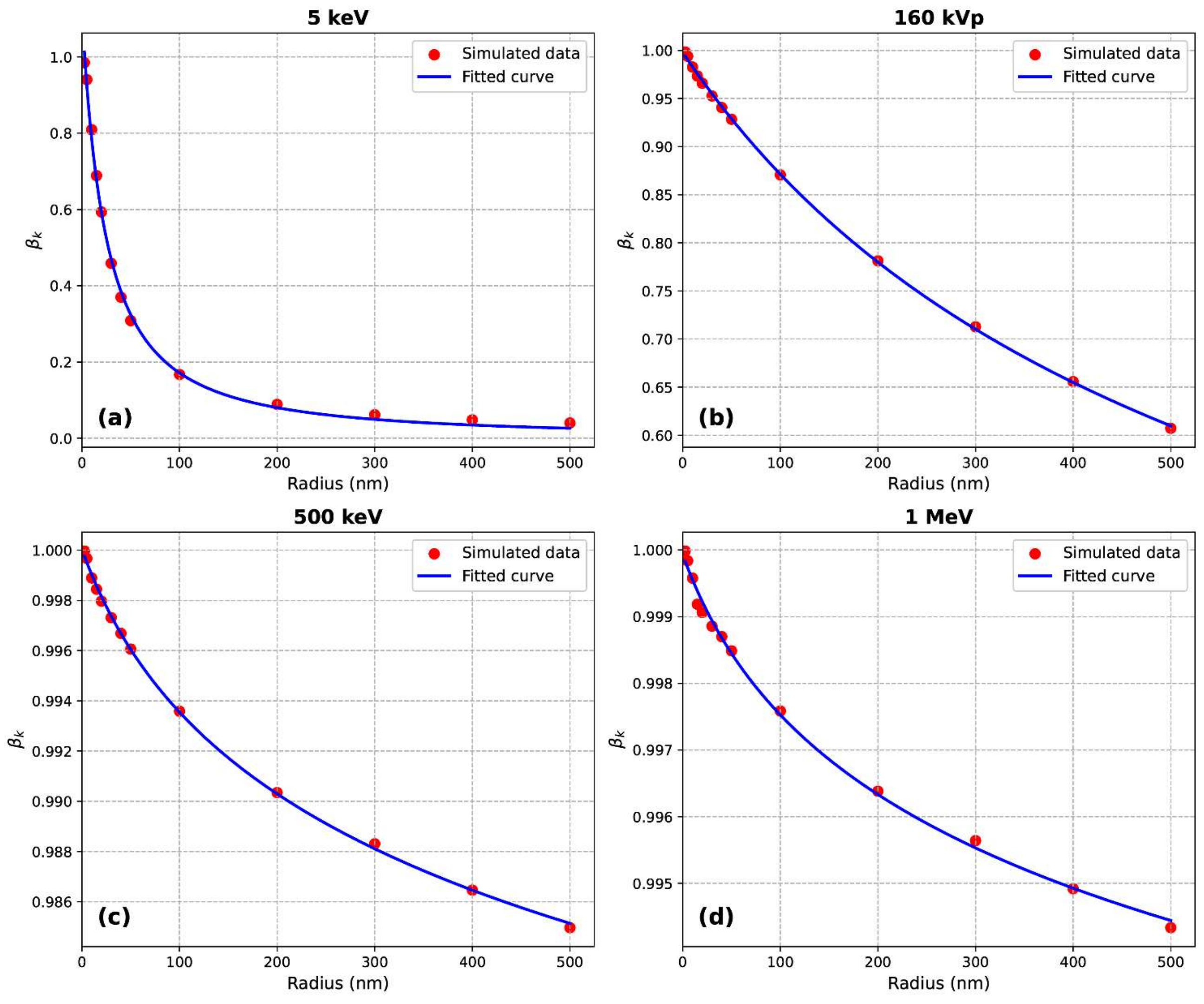

**Figure A1**. Simulated data points and fitted curves for the radius-dependence of gold-sphere energy-emission ratio $\beta_k$ under different incident photon energies.

## Appendix B. Calculation of total effective area and energy-emission ratio of nanoparticle systems

### B.1. Generation of randomly agglomerated nanoparticle models

Models of random agglomerates were generated via three-dimensional Diffusion-Limited Aggregation (DLA) simulations implemented in Python. Aglomorate growth was initiated from a seed nanoparticle of diameter $d$, radius $r$=$d$/2, positioned at the coordinate origin. The simulation domain was a cube centred at the origin with side length $L=20d$, which served as the boundary for nanoparticle random walks. Subsequently, a new nanoparticle was randomly created within the simulation domain and performed a random walk along six discrete directions $(\pm x, \pm y, \pm z)$. The nanoparticle was integrated into the existing agglomerate when its distance to the cluster fell below the sticking threshold set to $1.1d$, i.e., the centre-to-centre distance between two nanoparticles was slightly larger than the nanoparticle diameter. The central coordinates of the newly attached nanoparticle were then recorded. If the nanoparticle wandered out of the simulation domain, it was discarded and a new nanoparticle was created instead. This procedure was repeated until the agglomerate reached the predefined number of nanoparticles. Finally, the central coordinates $(x_i, y_i, z_i)$ of all constituent nanoparticles were exported and used as input to TOPAS to construct geometries for the subsequent Monte-Carlo simulation of random nanoparticle agglomerates. Within TOPAS, spherical nanoparticles of diameter $d$ were placed at each position according to the imported coordinates.

### B.2. Calculation of total energy-emission ratio

The medium surrounding the nanoparticle agglomerate was set to vacuum. The incident parallel photon beam was located $10d$ away from the origin, with a $20d$ diameter to fully cover the agglomerate. The total energy transfer $E_k$ (equation (29)) mediated by all the gold nanoparticles in the agglomerate was tallied using the custom volume scorer "ScoreKineticEnergy" (section 2.4.3). This scorer was defined on a spherical volume centered at the origin, serving as the parent component for all nanoparticles, with a diameter of $20d$ to enclose the entire agglomerate. A phase-space scorer was defined on the outer surface of this spherical volume to record the total kinetic energy of electrons escaping from the agglomerate. The TOPAS physics module and settings matched those adopted in the calibration for single-sphere energy-emission ratio (section 2.4.3).

### B.3. Calculation of total effective surface area

- Per-nanoparticle occlusion area for two-sphere geometry

Figure B1a shows two spheres $i$ and $j$ with the same radius $r$. The center-to-center distance is $d=|O_iO_j|>2r$. The occlusion region of sphere $j$ on sphere $i$ is defined as follows: two external tangent lines are drawn from the center $O_i$ of sphere $i$ to the surface of sphere $j$, with tangent points $P_1$ and $P_2$. These tangent lines intersect the surface of sphere $i$ at $Q_1$ and $Q_2$. The spherical cap region on the surface of sphere $i$, bounded by $Q_1$ and $Q_2$, constitutes the occlusion region of sphere $j$ on sphere $i$. Physically, the spherical cap marks the geometric angular range on sphere $i$ with direct line-of-sight to sphere $j$. Secondary electrons emitted into this angular range may encounter sphere $j$. Note that this definition does not rule out re-incidence of electrons emitted from other unoccluded surface regions (see Discussion).

In the right-angled triangle $O_iP_1O_j$,

$$|O_iP_1| = \sqrt{d^2 - r^2} \tag{B1}$$

The half-apex angle $\alpha$ of the spherical cap satisfies

$$\sin\alpha = \frac{r}{d}, \quad \cos\alpha = \frac{\sqrt{d^2 - r^2}}{d} \tag{B2}$$

Height of the spherical cap

$$h = r - r\cos\alpha = r(1 - \cos\alpha) \tag{B3}$$

The area of the spherical cap is defined as the occlusion area

$$S_{ij} = 2\pi rh = 2\pi r^2(1 - \cos\alpha) \tag{B4}$$

This formula satisfies the limiting cases: as the center-to-center distance $d \to \infty$, the occlusion area $S_{\mathrm{occ}} \to 0$; when $d = 2r$, $S_{\mathrm{occ}} = \pi r^2$, corresponding to one-quarter of the spherical surface being occluded.

- Per-nanoparticle occlusion area in multi-sphere agglomerates

When sphere $i$ is occluded by multiple spheres $j_1, j_2, \dots, j_n$, the total occluded region is the union of multiple spherical-cap regions. The area of this union can be calculated using the inclusion-exclusion principle

$$S_{\mathrm{union}} = \sum S_{ij1} - \sum S_{i,j1\cap j2} + \sum S_{i,j1\cap j2\cap j3} - \cdots + (-1)^{n+1} S_{i,j1\cap\cdots\cap jn} \tag{B5}$$

where $S_{i,j_1\cap j_2}$ denotes the overlapping area of the spherical caps induced by $j_1$ and $j_2$ on sphere $I$, and $S_{i,j1\cap\cdots\cap jn}$ denotes the overlapping area of the spherical caps generated by spheres $j_1, j_2, \dots, j_n$ on sphere $i$.

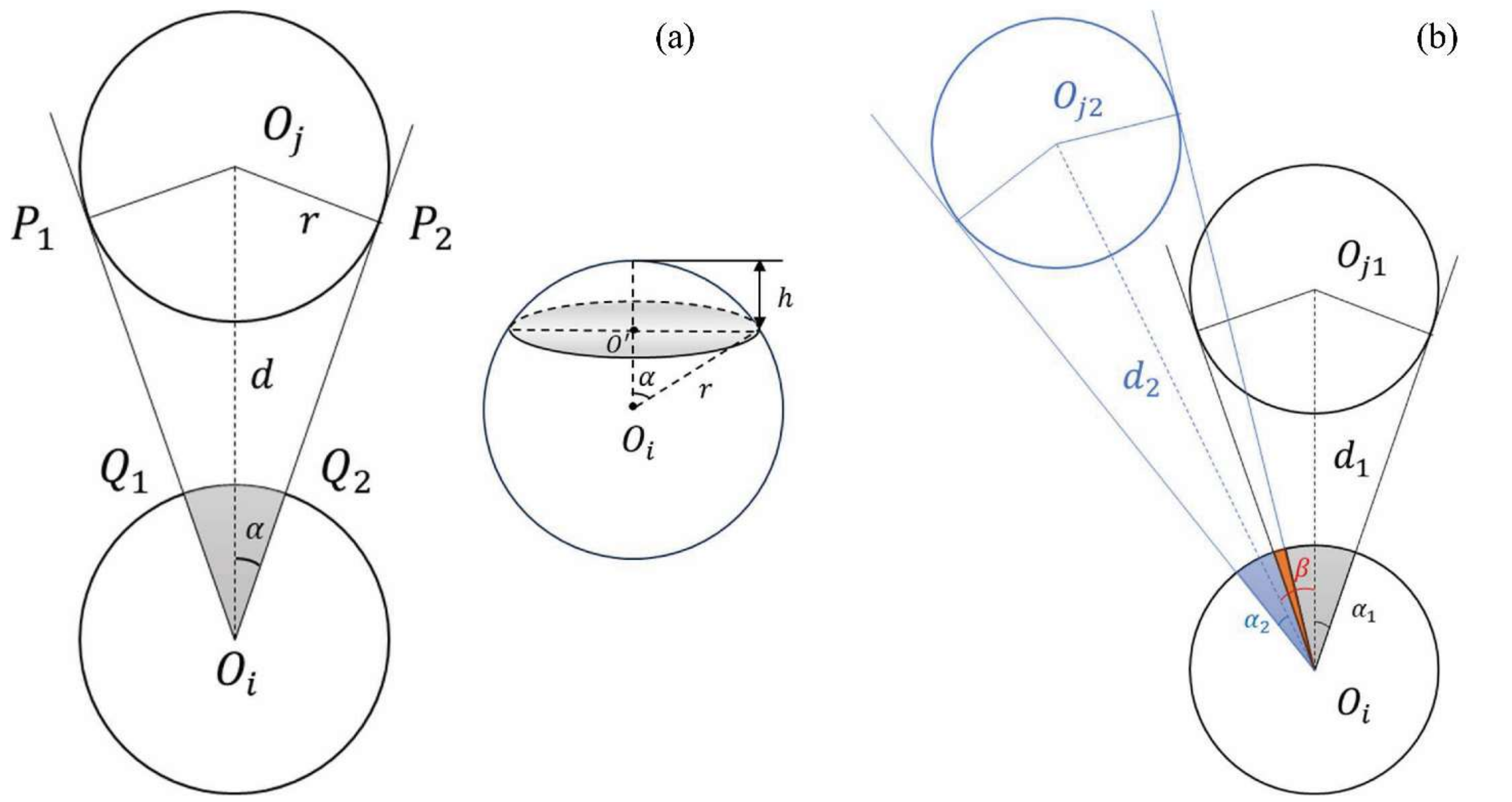

**Figure B1.** Schematic illustrating occlusion regions generated on a target sphere: (a) two-sphere configuration; (b) target sphere occluded by two neighbouring spheres within multi-sphere agglomerates.

- Monte-Carlo estimation for per-nanoparticle occlusion area

For a large number of occlusion spherical caps with complex relative positions on a sphere, rigorous analytical computation is infeasible. The Monte Carlo method is therefore employed to estimate the intersection areas, approximated by the fraction of abundant random sampling points falling inside the target region.

Consider a set of spherical caps $C=\{C_i\}_{i=1}^{n}$ on a unit sphere $\mathbb{S}^2$. Each spherical cap $C_i$ is defined by its centre $C_i(\theta_i, \varphi_i)$ and angular radius $\alpha_i$. The region of spherical cap $C_i$ reads $C_i = \{P(\theta, \varphi) \in \mathbb{S}^2 : d_{gc}(P, C_i) \leq \alpha_i\}$, where $\theta \in [0, \pi]$ denotes the polar angle and $\varphi \in [0, 2\pi)$ is the azimuthal angle. $d_{gc}(P, C_i)$ is the great-circle distance between point $P$ on the sphere and the centre $C_i$ of the spherical cap. On the unit sphere, this distance equals the central angle (in radians) subtended by the great-circle arc connecting the two points, which can be evaluated via the spherical law of cosines:
$d_{gc}=\arccos(\sin\theta \sin\theta_i \cos(\varphi - \varphi_i) + \cos\theta \cos\theta_i)$. The intersection region of all spherical caps is defined as:
$\Gamma = \bigcap_{i=1}^{n} C_i$. The area of the intersection region $\Gamma$ is denoted as $A(\Gamma)$.

The total surface area of the unit sphere is $A_{\text{sphere}}=4\pi$. From the principles of geometric probability, suppose $N$ points are uniformly and randomly sampled over the sphere, among which $M$ points fall inside the intersection region $\Gamma$. As $N \to \infty$, we have $M/N \to A(\Gamma)/A_{\text{sphere}}$. Accordingly, the area estimator is obtained as $\hat{A}(\Gamma)=A_{\text{sphere}}\, M/N = 4\pi\, M/N$, where $\hat{A}(\Gamma)$ represents the Monte Carlo estimate of $A(\Gamma)$.

## Appendix C. CPK-II: Simplified formulation and interpolation prediction scheme

Based on the explicit expression of the CPK-II model in equation (55), the independent parameter set $(K_m m_0, A, k, t)$ can be identified using a grid-search strategy over $t$. For each fixed candidate $t$, constrained nonlinear least-squares fitting can be performed to obtain optimal estimates of $(K_m m_0, A, k)$. The optimal $t$ and the corresponding parameter set $(K_m m_0, A, k, t)$ can be determined by scanning across all candidate $t$ values.

However, rigorous nonlinear least-squares fitting using equation (55) encounters numerical implementation difficulties due to the strong implicit coupling between $K_m m_0$ and $A$. The parameter $A$ appears both inside the nested power bracket and as an outer multiplier, giving rise to pronounced parameter degeneracy. Distinct parameter sets can produce nearly identical output curves, making the optimization highly sensitive to initial guesses. Since all unknowns enter nonlinearly, analytical solutions for the amplitude coefficients are unavailable for fixed $k$ and $t$. The fitting routine thus tends to converge to local minima and exhibits poor numerical stability. For this reason, further simplifications of equation (55), or specifically of the expression $\beta_k(R; E_\gamma)$ therein, are required for robust parameter fitting.

By applying a zero-order approximation (large-$R$ approximation for $R \gg d$) to the expression of $\beta_k(R; E_\gamma)$ in equation (32), we obtain a power-function

$$\beta_k(R; E_\gamma) = a(R + d)^b \approx aR^b \tag{C1}$$

Substituting equation (C1) into equation (37) yields a simplified three-parameter expression

$$\frac{\alpha''}{\alpha} = K_\mathrm{m} m_0 a R_0^b A^{bt+1} (1 - e^{-kC})^{bt+1} = A_\mathrm{m} (1 - e^{-k}\ )^{bt+1} \tag{C2}$$

where $A_\mathrm{m} = K_\mathrm{m} m_0 a R_0^b A^{bt+1}$ can be treated as a lumped independent parameter. Consequently, the model reduces to a three-parameter form with adjustable parameters $A_\mathrm{m}$, $k$ and $t$. A grid-search strategy similar to the approach adopted for CPK-I is used for the parameter $t$. For each specified $t$ value, $A_\mathrm{m}$ and $k$ are determined by solving the corresponding simultaneous equations. The optimal $t$ value is identified via error evaluation, and the corresponding parameter set $(A_\mathrm{m}, k, t)$ is used for interpolation prediction within the experimental concentration range. Goodness-of-fit analysis for the optimal $t$ follows the same fitting procedure used to obtain the optimal $\eta$ value in the CPK-I model, as described in Appendix E.

It should be noted that the three-parameter expression given by equation (C2), based on the power-law approximation of $\beta_k(R; E_\gamma)$ (equation (C1)), has a limited range of validity. Figure C1 shows that for 160 kVp X-ray irradiation, the relative error of expression (C1) with respect to the original equation (32) remains below 10% only when the $\beta_k$-equivalent single-sphere radius of intracellular nanoparticle agglomerates satisfies $R$ >2500 nm. For the 30 nm GNPs used for the validation, this implies an intracellular uptake exceeding $4.6 \times 10^6$ GNPs per cell on average. However, for 30 nm GNPs at the upper treatment concentration of 2 μg/mL, the mean number of intracellular nanoparticles per cell $\bar{\lambda}$ typically lies in the range of 10000 to 40000, and dense agglomeration rarely occurs. Accordingly, the $\beta_k$-equivalent single-sphere radius $R$ is much smaller than 2500 nm. This means that the power-law approximation of $\beta_k(R; E_\gamma)$ is valid only for sufficiently high GNP concentrations that yield adequate uptake and intracellular agglomeration (before the onset of severe cytotoxicity). At lower concentrations where nanoparticle uptake and agglomeration lie far below the above thresholds, a more accurate expression for $\beta_k(R; E_\gamma)$ and more rigorous parameter-determination approaches should be adopted.

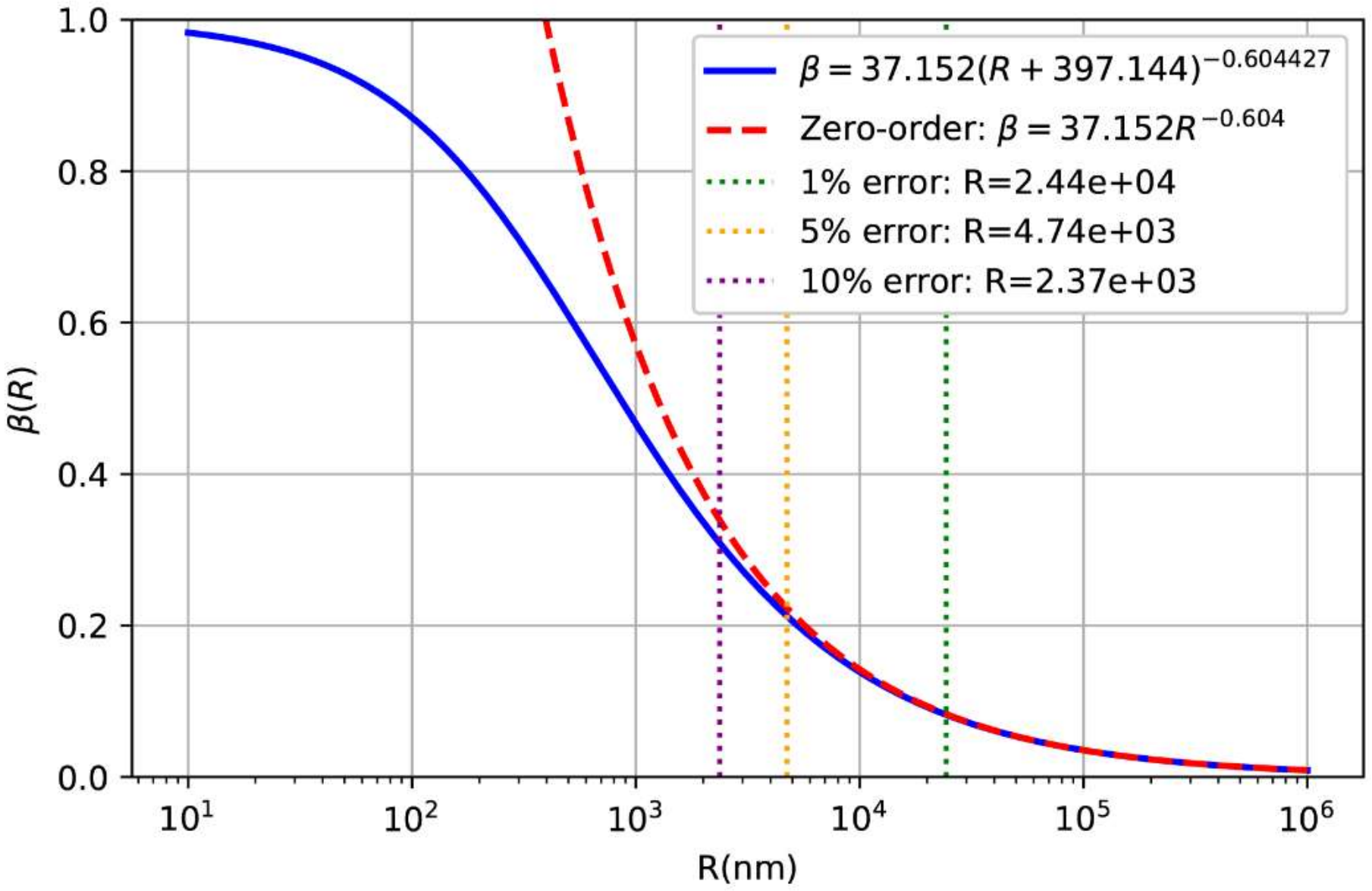

**Figure C1.** Comparison between the original fitted curve and its zero-order power-law approximation (large-R approximation for R>>d) of $\beta_k(R; E_\gamma)$ for 160 kVp X-ray irradiation. Blue solid line: $\beta_k = 37.152(R + 397.144)^{-0.604427)}$; red dashed line: $\beta_k = 37.152R^{-0.604}$. Vertical dotted lines indicate the radii corresponding to relative errors of 1%, 5%, and 10%, respectively.

## Appendix D. Cell experiments

### D.1. Cell culture

HepG2 cells (Shanghai Cell Bank of Chinese Academy of Sciences, China) were cultured in DMEM medium (Gibco, MD, USA) supplemented with 10% fetal bovine serum (Gemini Foundation, USA) and 1% Penicillin-streptomycin solution (Beyotime Biotechnology, China) in a humidified incubator at 37°C with 5% CO2.

### D.2. PEG-GNP cytotoxicity assay

The PEG-GNPs (Nanocs, USA) used for the CPK model validation are commercially available PEG-coated spherical gold nanoparticles (30 nm diameter) with a stock concentration of 0.1 mg/ml. Cytotoxicity assays were performed across a serial concentration gradient (4, 2, 1, 0.5, 0.25, 0.125 µg/ml). HepG2 cells were seeded in 96-well plates at a density of 5000 cells per well and divided into seven groups for co-incubation with PEG-GNPs at different concentrations. A 0 µg/ml group serves as the cell-only control, and a blank group containing pure medium (no cells, no GNPs) was included to substract background signals. Following an initial 24 h incubation, the culture medium was replaced with fresh medium containing PEG-GNPs at the seven tested concentrations. After 48 h of co-incubation, the medium was exchanged for DMEM supplemented with 10 % CCK-8 solution (Beyotime Biotechnology, China), and the plate was returned to the incubator for 0.5 h. Optical density (OD) was measured using a microplate rerader (MultiSkan FC, Thermo Scientific) at a detection wavelength of 450 nm and a reference wavelength of 650 nm. Relative cell viability (%) was calculated using $[(OD_{experimental} - OD_{blank})/(OD_{control} - OD_{blank})]\times 100\%$.

### D.3. Cell irradiation and clonogenic dose-survival assay

HepG2 cells were seeded in 60 mm dishes at varied seeding densities appropriate for each irradiation dose. Cells assigned to each irradiation-dose group were pre-treated with PEG-GNPs at different concentrations (0, 0.04, 0.2, 0.4, 2 µg/mL), with the 0 µg/mL dishes serving as the control group. Following 24h of co-incubation with PEG-GNPs, non-adherent cells were discard, and the culture medium was replaced. Cell irradiation was performed using a 160 kVp X-ray irradiator (IXS 1650, VJ Technologies Suzhou Inc, China) at doses of 0, 2, 4, 6, and 8 Gy, with a dose rate of 0.75 Gy/min. The irradiated cells were further incubated for 14 days to allow colony formation. Colonies were subsequently fixed with methanol and stained using 0.1% crystal violet solution (Beijing Solarbio Science & Technology, China). All experiments were repeated at least three times. Colonies containing at least 50 cells were counted under a microscope. Survival fraction at different doses were calculated from the plating efficiency of each group and presented as mean±SD. Data obtained at each GNP concentration were individually fitted to the LQ model, and the corresponding $(\alpha, \beta)$ parameter values were determined. Using experimental data from the control group and the boundary concentrations (0.04, 2 µg/ml), the relative values of $\alpha''/\alpha$ were computed for interpolation-based prediction during model validation.

## Appendix E. Interpolation-prediction of α″/α and goodness-of-fit evaluation for optimal η and t values

Interpolation prediction of the concentration-dependent $\alpha''/\alpha$ is performed from its functional expression using boundary $(\alpha''/\alpha, C)$ data. For the CPK-iso model, intermediate-concentration $\alpha''/\alpha$ values are predicted directly by equation (54) using the data at two boundary concentrations (high and low), whereby the fitted parameters ($A''$, $k$) can be solved explicitly. For the CPK-I and CPK-II models, a two-step procedure is adopted. First, optimal $\eta$ and $t$ values are identified via a grid-search strategy: candidate $\eta$ and $t$ values are varied, and equations (24) and (C2) are fitted against experimental $\alpha''/\alpha$ data from the intermediate validation concentrations to find the parameters yielding the highest goodness-of-fit. The resulting parameter sets $(A_s, k, \eta)$ and $(A_m, k, t)$ are then substituted into equations (24) and (C2), respectively, to generate interpolation-predicted $\alpha''/\alpha$ values at intermediate concentrations, with potentially improved accuracy.

When varying values of $\eta$ and $t$ are adopted to fit the LQ survival curves predicted by equation (24) (CPK-I) and equation (C2) (CPK-II) against clonogenic survival data at the intermediate validation concentrations, the goodness-of-fit between the predicted LQ survival and experimental survival data is quantified by the coefficient of determination

$$R^2 \equiv 1 - \frac{SS_{\text{res}}}{SS_{\text{tot}}}, \quad \left( SS_{\text{res}} = \sum_{i=1}^{n} (y_i - f_i)^2, \quad SS_{\text{tot}} = \sum_{i=1}^{n} (y_i - \bar{y})^2, \quad \bar{y} = \frac{1}{n}\sum_{i=1}^{n} y_i \right) \tag{E1}$$

where $n$ is the sample size (the total number of paired measured-predicted data points); $SS_{\text{res}}$ is the residual sum of squares; $SS_{\text{tot}}$ is the total sum of squares; $y_i$ denotes the experimentally measured survival fraction, and $f_i$ denotes the corresponding model-predicted survival value at different doses and concentrations; $\bar{y}$ is the mean of the experimentally measured survival fractions. In the present study, $n$ =10, corresponding to dose-survival data collected at five radiation doses (0, 2, 4, 6, 8 Gy) for the two intermediate validation concentrations (0.2, 0.4 µg/mL).

The $\eta$ and $t$ values yielding the highest $R^2$ were defined as optimal $\eta$ and $t$ parameters. In the present work, the $\eta$-dependent variation in $R^2$ for the CPK-I model and the $t$-depdenednt variation in $R^2$ for the CPK-II model are plotted in Fig E1a and Figure E1b, with the optimal values determined at $\eta \approx 0.92515$ and $t \approx 0.061980$, respectively.

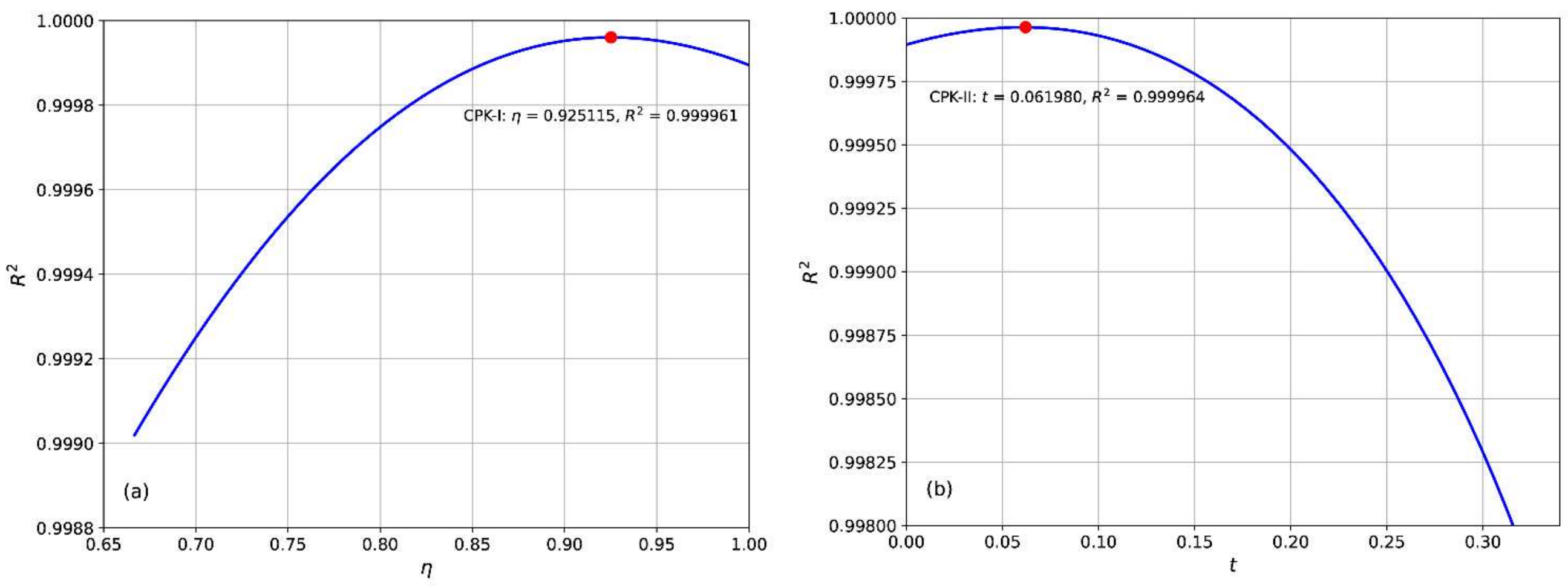

**Figure E1.** Goodness-of-fit for the agglomeration models (CPK-I and CPK-II) against experimental dose-survival data for different $\eta$ and $t$. The plots show the agglomeration-dependent variation in $R^2$ calculated for the LQ survival curves predicted under varied $\eta$ (CPK-I) and $t$ (CPK-II) values, identifying the optimal parameter values that yield the highest $R^2$. (a) Single-effect agglomeration model for surface-mediated biochemical enhancement (CPK-I model); (b) Single-effect agglomeration model for mass-energy-transfer-mediated loical-dose and radolysis enhancement (CPK-II model).

## Appendix F. Statistical metrics for overall prediction-error comparison

The root-mean-square deviation (RMSD), mean absolute deviation (MAD), standard deviation of prediction error (SDerr) and mean deviation (MeanDev) are defined respectively as

$$RMSD = \sqrt{\frac{1}{n}\sum_{i=1}^{n} {dev_i}^2}, \quad dev_i = \frac{P_i - E_i}{E_i} \times 100\% \tag{F1}$$

$$MAD = \frac{1}{n}\sum_{i=1}^{n} |dev_i|, \quad dev_i = \frac{P_i - E_i}{E_i} \times 100\% \tag{F2}$$

$$SDerr = \sqrt{\frac{1}{n-1}\sum_{i=1}^{n} \left(dev_i - \overline{dev}\right)^2}, \quad dev_i = \frac{P_i - E_i}{E_i} \times 100\%, \quad \overline{dev} = \frac{1}{n}\sum_{i=1}^{n} dev_i \tag{F3}$$

$$Mean\ Dev = \frac{1}{n}\sum_{i=1}^{n} dev_i, \quad dev_i = \frac{P_i - E_i}{E_i} \times 100\% \tag{F4}$$

where $P_i$ denotes the model-predicted cell survival fraction and $E_i$ denotes the corresponding experimentally measured survival fraction; $dev_i$ is the relative percentage deviation defined as $dev_i$=$(P_i - E_i/E_i)$×100%; $n$ is the sample size (the total number of paired experimental-predicted data points). For each separate intermediate validation concentration, $n$ =5, corresponding to the five radiation-dose points.